\documentclass[12pt,reqno]{amsart}%
\usepackage{graphicx}
\usepackage{amscd}
\usepackage{amsmath}
\usepackage{epsfig}
\usepackage{amsfonts}
\usepackage{amssymb}%
\providecommand{\U}[1]{\protect\rule{.1in}{.1in}}
\numberwithin{equation}{section}
\providecommand{\U}[1]{\protect\rule{.1in}{.1in}}
\providecommand{\U}[1]{\protect\rule{.1in}{.1in}}
\allowdisplaybreaks

\theoremstyle{plain}

\begin{document}
\title[Complex Electrodynamics]{Complex Form of Classical and Quantum Electrodynamics}
\dedicatory{\textquotedblleft Physical laws should have mathematical beauty.\negthinspace
\negthinspace\textquotedblright~---~P.~A.~M.~Dirac}\author{Sergey I. Kryuchkov}
\address{School of Mathematical and Statistical Sciences, Arizona State University,
Tempe, AZ 85287--1804, U.S.A.}
\email{sergeykryuchkov@yahoo.com}
\author{Nathan A. Lanfear}
\address{School of Mathematical and Statistical Sciences, Arizona State University,
Tempe, AZ 85287--1804, U.S.A.}
\email{nlanfear@asu.edu}
\author{Sergei~K.~Suslov}
\address{School of Mathematical and Statistical Sciences, Arizona State University,
Tempe, AZ 85287--1804, U.S.A.}
\email{sks@asu.edu}
\urladdr{http://hahn.la.asu.edu/\symbol{126}suslov/index.html}
\date{May 23, 2016}

\begin{abstract}
We consider a complex covariant form of the macroscopic Maxwell equations, in
a moving medium or at rest, following the original ideas of Minkowski. A
compact, Lorentz invariant, derivation of the energy-momentum tensor and the
corresponding differential balance equations are given. Conservation laws and
quantization of the electromagnetic field will be discussed in this covariant
approach elsewhere.

\end{abstract}
\maketitle

\section{Introduction}

Although a systematic study of electromagnetic phenomena in media is not
possible without methods of quantum mechanics, statistical physics and
kinetics, in practice a standard mathematical model based on phenomenological
Maxwell's equations provides a good approximation to many important problems.
As is well-known, one should be able to obtain the electromagnetic laws for
continuous media from those for the interaction of fields and point particles
\cite{Denisov89}, \cite{deGrootSuttorp}, \cite{Kaufman62}, \cite{LanLif8},
\cite{Lorentz16}, \cite{MinkowskiII}, \cite{ToptyginII}. As a result of the
hard work of several generations of researchers and engineers, the classical
electrodynamics, especially in its current complex covariant form, undoubtedly
satisfies Dirac's criteria of mathematical beauty\footnote{During a seminar at
Moscow State University on October~3, 1956, when asked to summarize his
philosophy of physics, Dirac wrote the above cited sentence on the blackboard
in capital letters \cite{Farmelo}, \cite{Jacob96}, \cite{TerRyb}.}, being a
state of the art mathematical description of nature.

In macroscopic electrodynamics, the volume (mechanical or ponderomotive)
forces, acting on a medium, and the corresponding energy density and energy
flux are introduced with the help of the energy-momentum tensors and
differential balance relations \cite{EinsteinLaub08}, \cite{Ginsburg89},
\cite{LanLif8}, \cite{Pauli}, \cite{Tamm}, \cite{ToptyginII}. These forces
occur in the equations of motion for a medium or individual charges and, in
principle, they can be experimentally tested \cite{GinsburgUgarovUFN76},
\cite{ObukhovHehl03}, \cite{Pfeiferetal07}, \cite{ToptyginLevinaUFN16} (see
also the references therein). But interpretation of the results should depend
on the accepted model of the interaction between the matter and radiation.

In this methodological note, we discuss a complex version of Minkowski's
phenomenological electrodynamics (at rest or in a moving medium) without
assuming any particular form of material equations as far as possible. Lorentz
invariance of the corresponding differential balance equations is emphasized
in view of long-standing uncertainties about the electromagnetic stresses and
momentum density, the so-called \textquotedblleft Abraham-Minkowski
controversy\textquotedblright\ (see, for example, \cite{Barnett10},
\cite{BolStolUFN75}, \cite{DodinFisch12}, \cite{Einstein16},
\cite{EinsteinLaub08}, \cite{GinsburgUFN73}, \cite{Ginsburg89},
\cite{GinsburgUgarovUFN76}, \cite{deGrootSuttorp}, \cite{IsraelSteward},
\cite{LanLif8}, \cite{MakRukUFN09}, \cite{MakRukUFN11},
\cite{Nesterenko2JMP16}, \cite{Obukhov08}, \cite{ObukhovHehl03}, \cite{Pauli},
\cite{PavlovUFN78}, \cite{Pfeiferetal07}, \cite{RikkenTigg12}, \cite{Skob73},
\cite{Tamm39}, \cite{Testa13}, \cite{ToptyginLevinaUFN16},
\cite{VeselagoUFN09}, \cite{VeselagoUFN11}, \cite{VeselagoShchavlevUFN10} and
the references therein).

The paper is organized as follows. In sections~2 to 4, we describe the
$3D$-complex version of Maxwell's equations and derive the corresponding
differential balance density laws for the electromagnetic fields. Their
covariant versions are given in sections~5 to 9. The case of a uniformly
moving medium is discussed in section~10 and complex Lagrangians are
introduced in section~11. Some useful tools are collected in appendices~A to C
for the reader's benefit.

\section{Maxwell's Equations in $3D$-Complex Form}

Traditionally, the macroscopic Maxwell equations in a fixed frame of reference
are given by%
\begin{equation}
\operatorname{curl}\mathbf{E}=-\frac{1}{c}\frac{\partial\mathbf{B}}{\partial
t}\ \left(  \text{Faraday}\right)  ,\qquad\qquad\operatorname{div}%
\mathbf{B}=0\ \left(  \text{no magnetic charge}\right)  \label{MaxwellHom}%
\end{equation}%
\begin{equation}
\operatorname{curl}\mathbf{H}=\frac{1}{c}\frac{\partial\mathbf{D}}{\partial
t}+\frac{4\pi}{c}\mathbf{j}_{\text{free}}\ \left(  \text{Biot}\&\text{Savart}%
\right)  ,\quad\operatorname{div}\mathbf{D}=4\pi\rho_{\text{free}}\ \left(
\text{Coulomb}\right)  \footnote{From this point, we shall write
$\rho_{\text{free}}=\rho$ and $\mathbf{j}_{\text{free}}=\mathbf{j}.$ A
detailed analysis of electromagnetic laws for continuous media from those for
point particles is given in \cite{deGrootSuttorp} (statistical description of
material media).}. \label{MaxwellNotHom}%
\end{equation}
Here, $\mathbf{E}$\ is the electric field, $\mathbf{D}$ is the displacement
field; $\mathbf{H}$\ is the magnetic field, $\mathbf{B}$ is the induction
field. These equations, which are obtained by averaging of microscopic
Maxwell's equations in the vacuum, provide a good mathematical description of
electromagnetic phenomena in various media, when complemented by the
corresponding material equations. In the simplest case of an isotropic medium
at rest, one usually has%
\begin{equation}
\mathbf{D}=\varepsilon\mathbf{E},\qquad\mathbf{B}=\mu\mathbf{H},\qquad
\mathbf{j}=\sigma\mathbf{E}, \label{MaterialConstant}%
\end{equation}
where $\varepsilon$ is the dielectric constant, $\mu$ is the magnetic
permeability, and $\sigma$ describes the conductivity of the medium (see, for
example, \cite{Abraham}, \cite{BarutEld}, \cite{Becker}, \cite{BolStolUFN75},
\cite{BolStolSb83}, \cite{Denisov89}, \cite{Einstein05},
\cite{EinsteinLaub08ab}, \cite{Fock65}, \cite{deGrootSuttorp},
\cite{Jackson2nd}, \cite{LanLif8}, \cite{Lorentz16}, \cite{PanoPhil62},
\cite{Pauli}, \cite{Stratton}, \cite{Tamm}, \cite{TerRyb}, \cite{ToptyginI},
\cite{ToptyginII} for fundamentals of classical electrodynamics).

Introduction of two complex fields%
\begin{equation}
\mathbf{F}=\mathbf{E}+i\mathbf{H},\qquad\mathbf{G}=\mathbf{D}+i\mathbf{B}
\label{FGcomplex}%
\end{equation}
allows one to rewrite the phenomenological Maxwell equations in the following
compact form%
\begin{equation}
\frac{i}{c}\left(  \frac{\partial\mathbf{G}}{\partial t}+4\pi\mathbf{j}%
\right)  =\operatorname{curl}\mathbf{F},\qquad\mathbf{j}=\mathbf{j}^{\ast},
\label{MaxwellComplexHom}%
\end{equation}%
\begin{equation}
\operatorname{div}\mathbf{G}=4\pi\rho,\qquad\rho=\rho^{\ast},
\label{MaxwellComplexNotHom}%
\end{equation}
where the asterisk stands for complex conjugation (see also \cite{BarutEld},
\cite{KrLanSus15} and \cite{Schwinger}). As we shall demonstrate, different
complex forms of Maxwell's equations are particularly convenient for study of
the corresponding \textquotedblleft energy-momentum\textquotedblright\ balance
equations for the electromagnetic fields in the presence of the
\textquotedblleft free\textquotedblright\ charges and currents in a medium.

\section{Hertz Symmetric Stress Tensor}

We begin from a complex $3D$-interpretation of the traditional symmetric
energy-momentum tensor \cite{Pauli}. By definition,%
\begin{align}
&  T_{pq}=\frac{1}{16\pi}\left[  F_{p}G_{q}^{\ast}+F_{p}^{\ast}G_{q}%
+F_{q}G_{p}^{\ast}+F_{q}^{\ast}G_{p}\right. \label{StressSymComplex}\\
&  \qquad\quad-\left.  \delta_{pq}\left(  \mathbf{F}\cdot\mathbf{G}^{\ast
}+\mathbf{F}^{\ast}\cdot\mathbf{G}\right)  \right]  =T_{qp}\qquad\left(
p,q=1,2,3\right) \nonumber
\end{align}
and the corresponding \textquotedblleft momentum\textquotedblright\ balance
equation,%
\begin{align}
&  \left(  \rho\mathbf{E}+\frac{1}{c}\mathbf{j}\times\mathbf{B}\right)
_{p}+\frac{\partial}{\partial t}\left[  \frac{1}{4\pi c}\left(  \mathbf{D}%
\times\mathbf{B}\right)  \right]  _{p}\label{MomentumBalance}\\
&  \ =\frac{\partial T_{pq}}{\partial x_{q}}+\frac{1}{16\pi}\left[
\operatorname{curl}\left(  \mathbf{F\times G}^{\ast}+\mathbf{F}^{\ast}%
\times\mathbf{G}\right)  \right]  _{p}\nonumber\\
&  \quad+\frac{1}{16\pi}\left(  F_{q}\frac{\partial G_{q}^{\ast}}{\partial
x_{p}}-G_{q}\frac{\partial F_{q}^{\ast}}{\partial x_{p}}+F_{q}^{\ast}%
\frac{\partial G_{q}}{\partial x_{p}}-G_{q}^{\ast}\frac{\partial F_{q}%
}{\partial x_{p}}\right)  ,\nonumber
\end{align}
can be obtained from Maxwell's equations (\ref{MaxwellComplexHom}%
)--(\ref{MaxwellComplexNotHom}) as a result of elementary but rather tedious
vector calculus calculations usually omitted in textbooks. (We use Einstein
summation convention over any two repeated indices unless otherwise stated. In
this paper, Greek indices run from $0$ to $3,$ while Latin indices may have
values from $1$ to $3$ inclusive.)

\noindent\textbf{Proof.} Indeed, in a $3D$-complex form,%
\begin{align}
&  \frac{\partial}{\partial x_{q}}\left(  F_{p}G_{q}^{\ast}+F_{q}G_{p}^{\ast
}-\delta_{pq}\mathbf{F}\cdot\mathbf{G}^{\ast}\right) \label{IdentityI}\\
&  \quad=\frac{\partial F_{p}}{\partial x_{q}}G_{q}^{\ast}+F_{p}\frac{\partial
G_{q}^{\ast}}{\partial x_{q}}+\frac{\partial F_{q}}{\partial x_{q}}G_{p}%
^{\ast}+F_{q}\frac{\partial G_{p}^{\ast}}{\partial x_{q}}-\frac{\partial
}{\partial x_{p}}\left(  F_{q}G_{q}^{\ast}\right) \nonumber\\
&  \quad=F_{q}\left(  \frac{\partial G_{p}^{\ast}}{\partial x_{q}}%
-\frac{\partial G_{q}^{\ast}}{\partial x_{p}}\right)  +\left(  \frac{\partial
F_{p}}{\partial x_{q}}-\frac{\partial F_{q}}{\partial x_{p}}\right)
G_{q}^{\ast}\nonumber\\
&  \quad\quad+F_{p}\operatorname{div}\mathbf{G}^{\ast}+G_{p}^{\ast
}\operatorname{div}\mathbf{F}\nonumber\\
&  \quad=F_{p}\operatorname{div}\mathbf{G}^{\ast}-\left(  \mathbf{F}%
\times\operatorname{curl}\mathbf{G}^{\ast}\right)  _{p}+G_{p}^{\ast
}\operatorname{div}\mathbf{F-}\left(  \mathbf{G}^{\ast}\times
\operatorname{curl}\mathbf{F}\right)  _{p}\nonumber
\end{align}
due to an identity \cite{Tamm}:%
\begin{equation}
\left(  \mathbf{A}\times\operatorname{curl}\mathbf{B}\right)  _{p}%
=A_{q}\left(  \frac{\partial B_{q}}{\partial x_{p}}-\frac{\partial B_{p}%
}{\partial x_{q}}\right)  . \label{AcrossBcurl}%
\end{equation}
Taking into account the complex conjugate, we derive%
\begin{align}
\frac{1}{2}  &  \frac{\partial}{\partial x_{q}}\left[  F_{p}G_{q}^{\ast}%
+F_{p}^{\ast}G_{q}+F_{q}G_{p}^{\ast}+F_{q}^{\ast}G_{p}-\delta_{pq}\left(
\mathbf{F}\cdot\mathbf{G}^{\ast}+\mathbf{F}^{\ast}\cdot\mathbf{G}\right)
\right] \label{FactI}\\
&  \quad=\frac{1}{2}\left(  \mathbf{F}\operatorname{div}\mathbf{G}^{\ast
}-\mathbf{G}^{\ast}\times\operatorname{curl}\mathbf{F}+\mathbf{F}^{\ast
}\operatorname{div}\mathbf{G}-\mathbf{G}\times\operatorname{curl}%
\mathbf{F}^{\ast}\right)  _{p}\nonumber\\
&  \qquad+\frac{1}{2}\left(  \mathbf{G}\operatorname{div}\mathbf{F}^{\ast
}-\mathbf{F}^{\ast}\times\operatorname{curl}\mathbf{G}+\mathbf{G}^{\ast
}\operatorname{div}\mathbf{F}-\mathbf{F}\times\operatorname{curl}%
\mathbf{G}^{\ast}\right)  _{p}\nonumber
\end{align}
as our first important fact.

On the other hand, in view of Maxwell's equations (\ref{MaxwellComplexHom}%
)--(\ref{MaxwellComplexNotHom}), one gets%
\begin{align}
&  \mathbf{F}\operatorname{div}\mathbf{G}^{\ast}-\mathbf{G}^{\ast}%
\times\operatorname{curl}\mathbf{F}\\
&  \quad=4\pi\rho\mathbf{F}+\frac{i}{c}\left(  \frac{\partial\mathbf{G}%
}{\partial t}\times\mathbf{G}^{\ast}+4\pi\mathbf{j\times G}^{\ast}\right)
\nonumber
\end{align}
and, with the help of its complex conjugate,%
\begin{align}
&  \mathbf{F}\operatorname{div}\mathbf{G}^{\ast}-\mathbf{G}^{\ast}%
\times\operatorname{curl}\mathbf{F}+\mathbf{F}^{\ast}\operatorname{div}%
\mathbf{G}-\mathbf{G}\times\operatorname{curl}\mathbf{F}^{\ast}\\
&  \quad=4\pi\rho\left(  \mathbf{F}+\mathbf{F}^{\ast}\right)  +\frac{i}%
{c}\frac{\partial}{\partial t}\left(  \mathbf{G}\times\mathbf{G}^{\ast
}\right)  +\frac{4\pi i}{c}\mathbf{j\times}\left(  \mathbf{G}^{\ast
}-\mathbf{G}\right)  ,\nonumber
\end{align}
or
\begin{align}
&  \frac{1}{2}\left(  \mathbf{F}\operatorname{div}\mathbf{G}^{\ast}%
-\mathbf{G}^{\ast}\times\operatorname{curl}\mathbf{F}+\mathbf{F}^{\ast
}\operatorname{div}\mathbf{G}-\mathbf{G}\times\operatorname{curl}%
\mathbf{F}^{\ast}\right) \label{FactII}\\
&  \quad=4\pi\left(  \rho\mathbf{E}+\frac{1}{c}\mathbf{j}\times\mathbf{B}%
\right)  +\frac{1}{c}\frac{\partial}{\partial t}\left(  \mathbf{D}%
\times\mathbf{B}\right)  ,\nonumber
\end{align}
providing the second important fact. (Up to the constant, the first term in
the right-hand side represents the density of Lorentz's force acting on the
\textquotedblleft free\textquotedblright\ charges and currents in the medium
under consideration \cite{Tamm39}, \cite{Tamm}.)

In view of (\ref{FactII}) and (\ref{FactI}), we can write%
\begin{align}
&  4\pi\left(  \rho\mathbf{E}+\frac{1}{c}\mathbf{j}\times\mathbf{B}\right)
_{p}+\frac{1}{c}\frac{\partial}{\partial t}\left(  \mathbf{D}\times
\mathbf{B}\right)  _{p}\label{FactsI&II}\\
&  \quad=\frac{1}{2}\frac{\partial}{\partial x_{q}}\left[  F_{p}G_{q}^{\ast
}+F_{p}^{\ast}G_{q}+F_{q}G_{p}^{\ast}+F_{q}^{\ast}G_{p}-\delta_{pq}\left(
\mathbf{F}\cdot\mathbf{G}^{\ast}+\mathbf{F}^{\ast}\cdot\mathbf{G}\right)
\right] \nonumber\\
&  \quad\quad-\frac{1}{2}\left(  \mathbf{G}\operatorname{div}\mathbf{F}^{\ast
}-\mathbf{F}^{\ast}\times\operatorname{curl}\mathbf{G}+\mathbf{G}^{\ast
}\operatorname{div}\mathbf{F}-\mathbf{F}\times\operatorname{curl}%
\mathbf{G}^{\ast}\right)  _{p}\nonumber\\
&  \quad=\frac{1}{4}\frac{\partial}{\partial x_{q}}\left[  F_{p}G_{q}^{\ast
}+F_{p}^{\ast}G_{q}+F_{q}G_{p}^{\ast}+F_{q}^{\ast}G_{p}-\delta_{pq}\left(
\mathbf{F}\cdot\mathbf{G}^{\ast}+\mathbf{F}^{\ast}\cdot\mathbf{G}\right)
\right] \nonumber\\
&  \quad\quad-\frac{1}{4}\left(  \mathbf{G}\operatorname{div}\mathbf{F}^{\ast
}-\mathbf{F}^{\ast}\times\operatorname{curl}\mathbf{G}+\mathbf{G}^{\ast
}\operatorname{div}\mathbf{F}-\mathbf{F}\times\operatorname{curl}%
\mathbf{G}^{\ast}\right)  _{p}\nonumber\\
&  \quad\quad+\frac{1}{4}\left(  \mathbf{F}\operatorname{div}\mathbf{G}^{\ast
}-\mathbf{G}^{\ast}\times\operatorname{curl}\mathbf{F}+\mathbf{F}^{\ast
}\operatorname{div}\mathbf{G}-\mathbf{G}\times\operatorname{curl}%
\mathbf{F}^{\ast}\right)  _{p}\nonumber\\
&  \quad=4\pi\frac{\partial T_{pq}}{\partial x_{q}}+\frac{1}{4}\left(
\mathbf{F}\operatorname{div}\mathbf{G}^{\ast}-\mathbf{G}^{\ast}%
\operatorname{div}\mathbf{F}+\mathbf{F}^{\ast}\operatorname{div}%
\mathbf{G}-\mathbf{G}\operatorname{div}\mathbf{F}^{\ast}\right)
_{p}\nonumber\\
&  \quad\quad+\frac{1}{4}\left(  \mathbf{F}\times\operatorname{curl}%
\mathbf{G}^{\ast}-\mathbf{G}^{\ast}\times\operatorname{curl}\mathbf{F}%
+\mathbf{F}^{\ast}\times\operatorname{curl}\mathbf{G}-\mathbf{G}%
\times\operatorname{curl}\mathbf{F}^{\ast}\right)  _{p}.\nonumber
\end{align}
Finally, in the last two lines, one can utilize the following differential
vector calculus identity,%
\begin{align}
&  \left[  \mathbf{A}\operatorname{div}\mathbf{B}-\mathbf{B}\operatorname{div}%
\mathbf{A}+\mathbf{A}\times\operatorname{curl}\mathbf{B}-\mathbf{B}%
\times\operatorname{curl}\mathbf{A}-\operatorname{curl}\left(  \mathbf{A}%
\times\mathbf{B}\right)  \right]  _{p}\label{VectorIdentity}\\
&  \quad=A_{q}\frac{\partial B_{q}}{\partial x_{p}}-B_{q}\frac{\partial A_{q}%
}{\partial x_{p}},\nonumber
\end{align}
see (\ref{A5}), with $\mathbf{A}=\mathbf{F},$ $\mathbf{B}=\mathbf{G}^{\ast}$
and its complex conjugates, in order to obtain (\ref{MomentumBalance}) and/or
(\ref{MomentumBalancedReal}), which completes the proof. (An independent proof
will be given in section~7.)

Derivation of the corresponding differential \textquotedblleft
energy\textquotedblright\ balance equation is much simpler. By
(\ref{MaxwellComplexHom}),%
\begin{equation}
\mathbf{F}\cdot\frac{\partial\mathbf{G}^{\ast}}{\partial t}+\mathbf{F}^{\ast
}\cdot\frac{\partial\mathbf{G}}{\partial t}+4\pi\mathbf{j}\cdot\left(
\mathbf{F}+\mathbf{F}^{\ast}\right)  =\frac{c}{i}\operatorname{div}\left(
\mathbf{F}\times\mathbf{F}^{\ast}\right)
\end{equation}
due to a familiar vector calculus identity (\ref{A1}):
\begin{equation}
\operatorname{div}\left(  \mathbf{A}\times\mathbf{B}\right)  =\mathbf{B}%
\cdot\operatorname{curl}\mathbf{A}-\mathbf{A}\cdot\operatorname{curl}%
\mathbf{B}.
\end{equation}
In a traditional form,%
\begin{equation}
\frac{1}{4\pi}\left(  \mathbf{E}\cdot\frac{\partial\mathbf{D}}{\partial
t}+\mathbf{H}\cdot\frac{\partial\mathbf{B}}{\partial t}\right)  +\mathbf{j}%
\cdot\mathbf{E}+\operatorname{div}\left(  \frac{c}{4\pi}\mathbf{E}%
\times\mathbf{H}\right)  =0
\end{equation}
(see, for example, \cite{Denisov89}, \cite{Tamm}), where one can substitute%
\begin{align}
&  \mathbf{E}\cdot\frac{\partial\mathbf{D}}{\partial t}+\mathbf{H}\cdot
\frac{\partial\mathbf{B}}{\partial t}=\frac{1}{2}\frac{\partial}{\partial
t}\left(  \mathbf{E}\cdot\mathbf{D}+\mathbf{H}\cdot\mathbf{B}\right) \\
&  \quad+\frac{1}{2}\left(  \mathbf{E}\cdot\frac{\partial\mathbf{D}}{\partial
t}-\mathbf{D}\cdot\frac{\partial\mathbf{E}}{\partial t}+\mathbf{H}\cdot
\frac{\partial\mathbf{B}}{\partial t}-\mathbf{B}\cdot\frac{\partial\mathbf{H}%
}{\partial t}\right)  .\nonumber
\end{align}

As a result, $3D$-differential \textquotedblleft
energy-momentum\textquotedblright\ balance equations are given by%
\begin{align}
&  \frac{\partial}{\partial t}\left(  \frac{\mathbf{E}\cdot\mathbf{D}%
+\mathbf{H}\cdot\mathbf{B}}{8\pi}\right)  +\operatorname{div}\left(  \frac
{c}{4\pi}\mathbf{E}\times\mathbf{H}\right)  +\mathbf{j}\cdot\mathbf{E}%
\label{EnergyBalance}\\
&  \ +\frac{1}{8\pi}\left(  \mathbf{E}\cdot\frac{\partial\mathbf{D}}{\partial
t}-\mathbf{D}\cdot\frac{\partial\mathbf{E}}{\partial t}+\mathbf{H}\cdot
\frac{\partial\mathbf{B}}{\partial t}-\mathbf{B}\cdot\frac{\partial\mathbf{H}%
}{\partial t}\right)  =0\nonumber
\end{align}
and%
\begin{align}
&  -\frac{\partial}{\partial t}\left[  \frac{1}{4\pi c}\left(  \mathbf{D}%
\times\mathbf{B}\right)  \right]  _{p}+\frac{\partial T_{pq}}{\partial x_{q}%
}-\left(  \rho\mathbf{E}+\frac{1}{c}\mathbf{j}\times\mathbf{B}\right)
_{p}\label{MomentumBalancedReal}\\
&  \ +\frac{1}{8\pi}\left[  \operatorname{curl}\left(  \mathbf{E\times
D}+\mathbf{H}\times\mathbf{B}\right)  \right]  _{p}\nonumber\\
&  \quad+\frac{1}{8\pi}\left(  \mathbf{E}\cdot\frac{\partial\mathbf{D}%
}{\partial x_{p}}-\mathbf{D}\cdot\frac{\partial\mathbf{E}}{\partial x_{p}%
}+\mathbf{H}\cdot\frac{\partial\mathbf{B}}{\partial x_{p}}-\mathbf{B}%
\cdot\frac{\partial\mathbf{H}}{\partial x_{p}}\right)  =0,\nonumber
\end{align}
respectively (see also \cite{GinsburgUgarovUFN76}, \cite{MakRukUFN09}). The
real form of the symmetric stress tensor (\ref{StressSymComplex}), namely,%
\begin{align}
&  T_{pq}=\frac{1}{8\pi}\left[  E_{p}D_{q}+E_{q}D_{p}+H_{p}B_{q}+H_{q}%
B_{p}\right. \label{StressSymReal}\\
&  \qquad\qquad\left.  -\delta_{pq}\left(  \mathbf{E}\cdot\mathbf{D}%
+\mathbf{H}\cdot\mathbf{B}\right)  \right]  \qquad\left(  p,q=1,2,3\right)
,\nonumber
\end{align}
is due to Hertz \cite{Pauli}.

Equations (\ref{EnergyBalance})--(\ref{MomentumBalancedReal}) are related to a
fundamental concept of conservation of mechanical and electromagnetic energy
and momentum. Here, these balance conditions are presented in differential
forms in terms of the corresponding local field densities. They can be
integrated over a given volume in $\left.
\mathbb{R}
\right.  ^{3}$ in order to obtain, in a traditional way, the corresponding
conservation laws of the electromagnetic fields (see, for example,
\cite{LanLif2}, \cite{LanLif8}, \cite{TerRyb}, \cite{ToptyginI},
\cite{ToptyginII}). These laws made it necessary to ascribe a definite linear
momentum and energy to the field of an electromagnetic wave, which can be
observed, for example, as light pressure.

\noindent\textbf{Note.} At this point, the Lorentz invariance of these
differential balance equations is not obvious in our $3D$-analysis. But one
can introduce the four-vector $x^{\mu}=\left(  ct,\mathbf{r}\right)  $ and try
to match (\ref{EnergyBalance})--(\ref{MomentumBalancedReal}) with the
expression,%
\begin{equation}
\frac{\partial}{\partial x^{\nu}}T_{\mu}^{\ \nu}=\frac{\partial T_{\mu}^{\ 0}%
}{\partial x_{0}}+\frac{\partial T_{\mu}^{\ q}}{\partial x_{q}}\qquad\left(
\mu,\nu=0,1,2,3;\quad p,q=1,2,3\right)  , \label{4TensorDiff}%
\end{equation}
as an initial step, in order to guess the corresponding four-tensor form. An
independent covariant derivation will be given in section~7.

\noindent\textbf{Note.} In an isotropic nonhomogeneous variable medium
(without dispersion and/or compression), when $\mathbf{D}=\varepsilon\left(
\mathbf{r},t\right)  \mathbf{E}$ and $\mathbf{B}=\mu\left(  \mathbf{r}%
,t\right)  \mathbf{H},$ the \textquotedblleft ponderomotive
forces\textquotedblright\ in (\ref{EnergyBalance}) and
(\ref{MomentumBalancedReal}) take the form \cite{Tamm}:%
\begin{align}
&  \mathbf{E}\cdot\frac{\partial\mathbf{D}}{\partial x^{\nu}}-\mathbf{D}%
\cdot\frac{\partial\mathbf{E}}{\partial x^{\nu}}+\mathbf{H}\cdot\frac
{\partial\mathbf{B}}{\partial x^{\nu}}-\mathbf{B}\cdot\frac{\partial
\mathbf{H}}{\partial x^{\nu}}\label{PForce}\\
&  \quad=\dfrac{\partial\varepsilon}{\partial x^{\nu}}\mathbf{E}^{2}%
+\dfrac{\partial\mu}{\partial x^{\nu}}\mathbf{H}^{2}=\left(
\begin{array}
[c]{c}%
\dfrac{1}{c}\left(  \dfrac{\partial\varepsilon}{\partial t}\mathbf{E}%
^{2}+\dfrac{\partial\mu}{\partial t}\mathbf{H}^{2}\right)  \bigskip\\
\mathbf{E}^{2}\nabla\varepsilon+\mathbf{H}^{2}\nabla\mu
\end{array}
\right)  ,\nonumber
\end{align}
which may be interpreted as a four-vector \textquotedblleft
energy-force\textquotedblright\ acting from an inhomogeneous and time-variable
medium. Its covariance is analyzed in section~7.

\section{\textquotedblleft Angular Momentum\textquotedblright\ Balance}

The $3D$-\textquotedblleft linear momentum\textquotedblright\ differential
balance equation (\ref{MomentumBalancedReal}), can be rewritten in a more
compact form,%
\begin{equation}
\frac{\partial T_{pq}}{\partial x_{q}}=\mathcal{F}_{p}+\frac{\partial
\mathcal{G}_{p}}{\partial t},\qquad\overrightarrow{\mathcal{G}}=\frac{1}{4\pi
c}\left(  \mathbf{D}\times\mathbf{B}\right)  ,\label{CompactMomentumBalance}%
\end{equation}
with the help of the Hertz symmetric stress tensor $T_{pq}=T_{qp}$ defined by
(\ref{StressSymReal}). A \textquotedblleft net force\textquotedblright\ is
given by%
\begin{align}
\mathcal{F}_{p} &  =\left(  \rho\mathbf{E}+\frac{1}{c}\mathbf{j}%
\times\mathbf{B}\right)  _{p}-\frac{1}{8\pi}\left[  \operatorname{curl}\left(
\mathbf{E\times D}+\mathbf{H}\times\mathbf{B}\right)  \right]  _{p}%
\label{NetForce}\\
&  -\frac{1}{8\pi}\left(  \mathbf{E}\cdot\frac{\partial\mathbf{D}}{\partial
x_{p}}-\mathbf{D}\cdot\frac{\partial\mathbf{E}}{\partial x_{p}}+\mathbf{H}%
\cdot\frac{\partial\mathbf{B}}{\partial x_{p}}-\mathbf{B}\cdot\frac
{\partial\mathbf{H}}{\partial x_{p}}\right)  .\nonumber
\end{align}
In this notation, we state the  $3D$-\textquotedblleft angular
momentum\textquotedblright\ differential balance equation as follows%
\begin{equation}
\frac{\partial M_{pq}}{\partial x_{q}}=\mathcal{T}_{p}+\frac{\partial
\mathcal{L}_{p}}{\partial t},\qquad\overrightarrow{\mathcal{L}}%
=\mathbf{r\times}\overrightarrow{\mathcal{G}},\quad\overrightarrow{\mathcal{T}%
}=\mathbf{r\times}\overrightarrow{\mathcal{F}},\label{VectorMomentumBalance}%
\end{equation}
where the \textquotedblleft field angular momentum density\textquotedblright%
\ is defined by%
\begin{equation}
\overrightarrow{\mathcal{L}}=\frac{1}{4\pi c}\mathbf{r\times}\left(
\mathbf{D}\times\mathbf{B}\right)  \label{FieldAngularMomentumDensity}%
\end{equation}
and the \textquotedblleft flux of angular momentum\textquotedblright\ is
described by the following tensor \cite{Jackson2nd}:%
\begin{equation}
M_{pq}=e_{prs}x_{r}T_{sq}.\label{TensorMomentum}%
\end{equation}
(Here, $e_{pqr}$ is the totally anti-symmetric Levi-Civita symbol with
$e_{123}=+1$). An elementary example of conservation of the total angular
momentum is discussed in \cite{Tamm}.

\noindent\textbf{Proof.} Indeed, in view of (\ref{CompactMomentumBalance}),
one can write%
\begin{align}
\frac{\partial M_{pq}}{\partial x_{q}}  &  =e_{prs}T_{sr}+e_{prs}x_{r}%
\frac{\partial T_{sq}}{\partial x_{q}}\label{AngularMomentumProof}\\
&  =e_{pqr}x_{q}\mathcal{F}_{r}+\frac{\partial}{\partial t}\left(
e_{pqr}x_{q}\mathcal{G}_{r}\right)  ,\nonumber
\end{align}
which completes the proof.

\noindent\textbf{Note.} Once again, in $3D$-form, the Lorentz invariance of
this differential balance equation for the local densities is not obvious. An
independent covariant derivation will be given in section~8.

\section{Complex Covariant Form of Macroscopic Maxwell's Equations}

With the help of complex fields $\mathbf{F}=\mathbf{E}+i\mathbf{H}$ and
$\mathbf{G}=\mathbf{D}+i\mathbf{B},$ we introduce the following anti-symmetric
four-tensor,%
\begin{equation}
Q^{\mu\nu}=-Q^{\nu\mu}=\left(
\begin{array}
[c]{cccc}%
0 & -G_{1} & -G_{2} & -G_{3}\\
G_{1} & 0 & iF_{3} & -iF_{2}\\
G_{2} & -iF_{3} & 0 & iF_{1}\\
G_{3} & iF_{2} & -iF_{1} & 0
\end{array}
\right)  \label{ComplexQTensor}%
\end{equation}
and use the standard four-vectors, $x^{\mu}=\left(  ct,\mathbf{r}\right)  $
and $j^{\mu}=\left(  c\rho,\mathbf{j}\right)  $ for contravariant coordinates
and current, respectively.

Maxwell's equations then take the covariant form \cite{KrLanSus15},
\cite{LapUh31}:
\begin{equation}
\frac{\partial}{\partial x^{\nu}}Q^{\mu\nu}=-\frac{\partial}{\partial x^{\nu}%
}Q^{\nu\mu}=-\frac{4\pi}{c}j^{\mu} \label{CovariantMaxwell}%
\end{equation}
with summation over two repeated indices. Indeed, in block form, we have%
\begin{equation}
\frac{\partial Q^{\mu\nu}}{\partial x^{\nu}}=\frac{\partial}{\partial x^{\nu}%
}\left(
\begin{array}
[c]{cc}%
0 & -G_{q}\medskip\\
G_{p} & ie_{pqr}F_{r}%
\end{array}
\right)  =\left(
\begin{array}
[c]{c}%
-\operatorname{div}\mathbf{G}=-4\pi\rho\medskip\\
\dfrac{1}{c}\dfrac{\partial\mathbf{G}}{\partial t}+i\operatorname{curl}%
\mathbf{F}=-\dfrac{4\pi}{c}\mathbf{j}%
\end{array}
\right)  , \label{CMProof}%
\end{equation}
which verifies this fact. The continuity equation,%
\begin{equation}
0\equiv\frac{\partial^{2}Q^{\mu\nu}}{\partial x^{\mu}\partial x^{\nu}}%
=-\frac{4\pi}{c}\frac{\partial j^{\mu}}{\partial x^{\mu}},
\label{CovariantContinuity}%
\end{equation}
or in the $3D$-form,%
\begin{equation}
\frac{\partial\rho}{\partial t}+\operatorname{div}\mathbf{j}=0,
\label{ContinuityEq}%
\end{equation}
describes conservation of the electrical charge. The latter equation can also
be derived in the complex $3D$-form from (\ref{MaxwellComplexHom}%
)--(\ref{MaxwellComplexNotHom}).

\noindent\textbf{Note.} In vacuum, when $\mathbf{G}=\mathbf{F}$ and $\rho=0,$
$\mathbf{j}=0,$ one can write due to (\ref{B1})--(\ref{B2}):%
\begin{equation}
Q^{\mu\nu}=F^{\mu\nu}-\frac{i}{2}e^{\mu\nu\sigma\tau}F_{\sigma\tau},\qquad
F^{\mu\nu}=g^{\mu\sigma}g^{\nu\tau}F_{\sigma\tau},\qquad g_{\mu\sigma}%
g_{\nu\tau}Q^{\sigma\tau}=Q_{\mu\nu}.\label{SelfDual}%
\end{equation}
As a result, the following self-duality property holds%
\begin{equation}
e_{\mu\nu\sigma\tau}Q^{\sigma\tau}=2iQ_{\mu\nu},\qquad2iQ^{\mu\nu}=e^{\mu
\nu\sigma\tau}Q_{\sigma\tau}\label{DualVacuum}%
\end{equation}
(see, for example, \cite{Bia:Bia75}, \cite{KrLanSus16} and appendix~B). Two
covariant forms of Maxwell's equations are given by%
\begin{equation}
\partial_{\nu}Q^{\mu\nu}=0,\qquad\partial^{\nu}Q_{\mu\nu}%
=0,\label{CovariantMaxwellVacuum}%
\end{equation}
where $\partial^{\nu}=g^{\nu\mu}\partial_{\mu},$ $\partial_{\mu}%
=\partial/\partial x^{\mu}$ and $g_{\mu\nu}=g^{\mu\nu}=$diag$\left(
1,-1,-1,-1\right)  .$ The last equation can be derived from a more general
equation, involving a rank three tensor,%
\begin{equation}
g^{\alpha\alpha}e_{\alpha\mu\nu\tau}\partial^{\nu}Q^{\tau\beta}-g^{\alpha
\alpha}e_{\beta\mu\nu\tau}\partial^{\nu}Q^{\tau\alpha}=-i\partial_{\mu
}Q^{\alpha\beta}\label{MaxwellPauliLubanskii}%
\end{equation}
($\alpha,\beta=0,1,2,3$ are fixed; no summation is assumed over these two
indices), which is related to the Pauli-Luba\'{n}ski vector from the
representation theory of the Poincar\'{e} group \cite{KrLanSus15}. Different
spinor forms of Maxwell's equations are analyzed in \cite{KrLanSus16} (see
also the references therein).

\section{Dual Electromagnetic Field Tensors}

Two dual anti-symmetric field tensors of complex fields, $\mathbf{F}%
=\mathbf{E}+i\mathbf{H}$ and $\mathbf{G}=\mathbf{D}+i\mathbf{B},$ are given
by
\begin{align}
Q^{\mu\nu}  &  =\left(
\begin{array}
[c]{cccc}%
0 & -G_{1} & -G_{2} & -G_{3}\\
G_{1} & 0 & iF_{3} & -iF_{2}\\
G_{2} & -iF_{3} & 0 & iF_{1}\\
G_{3} & iF_{2} & -iF_{1} & 0
\end{array}
\right)  =R^{\mu\nu}+iS^{\mu\nu}\label{ComplexQ}\\
&  =\left(
\begin{array}
[c]{cccc}%
0 & -D_{1} & -D_{2} & -D_{3}\\
D_{1} & 0 & -H_{3} & H_{2}\\
D_{2} & H_{3} & 0 & -H_{1}\\
D_{3} & -H_{2} & H_{1} & 0
\end{array}
\right)  +i\left(
\begin{array}
[c]{cccc}%
0 & -B_{1} & -B_{2} & -B_{3}\\
B_{1} & 0 & E_{3} & -E_{2}\\
B_{2} & -E_{3} & 0 & E_{1}\\
B_{3} & E_{2} & -E_{1} & 0
\end{array}
\right) \nonumber
\end{align}
and%
\begin{align}
P_{\mu\nu}  &  =\left(
\begin{array}
[c]{cccc}%
0 & F_{1} & F_{2} & F_{3}\\
-F_{1} & 0 & iG_{3} & -iG_{2}\\
-F_{2} & -iG_{3} & 0 & iG_{1}\\
-F_{3} & iG_{2} & -iG_{1} & 0
\end{array}
\right)  =F_{\mu\nu}+iG_{\mu\nu}\label{ComplexP}\\
&  =\left(
\begin{array}
[c]{cccc}%
0 & E_{1} & E_{2} & E_{3}\\
-E_{1} & 0 & -B_{3} & B_{2}\\
-E_{2} & B_{3} & 0 & -B_{1}\\
-E_{3} & -B_{2} & B_{1} & 0
\end{array}
\right)  +i\left(
\begin{array}
[c]{cccc}%
0 & H_{1} & H_{2} & H_{3}\\
-H_{1} & 0 & D_{3} & -D_{2}\\
-H_{2} & -D_{3} & 0 & D_{1}\\
-H_{3} & D_{2} & -D_{1} & 0
\end{array}
\right)  .\nonumber
\end{align}
The real part of the latter represents the standard electromagnetic field
tensor in a medium \cite{BarutEld}, \cite{Pauli}, \cite{ToptyginII}. As for
the imaginary part of (\ref{ComplexQ}), which, ironically, Pauli called an
\textquotedblleft artificiality\textquotedblright\ in view of its non-standard
behavior under spatial inversion \cite{Pauli}, the use of complex conjugation
restores this important symmetry for our complex field tensors.

The dual tensor identities are given by%
\begin{equation}
e_{\mu\nu\sigma\tau}Q^{\sigma\tau}=2iP_{\mu\nu},\qquad2iQ^{\mu\nu}=e^{\mu
\nu\sigma\tau}P_{\sigma\tau}. \label{PQdual}%
\end{equation}
Here $e^{\mu\nu\sigma\tau}=-e_{\mu\nu\sigma\tau}$ and $e_{0123}=+1$ is the
Levi-Civita four-symbol \cite{Fock64}. Then%
\begin{equation}
6i\frac{\partial Q^{\mu\nu}}{\partial x^{\nu}}=e^{\mu\nu\lambda\sigma}\left(
\frac{\partial P_{\lambda\sigma}}{\partial x^{\nu}}+\frac{\partial
P_{\nu\lambda}}{\partial x^{\sigma}}+\frac{\partial P_{\sigma\nu}}{\partial
x^{\lambda}}\right)  \label{QPIdentity}%
\end{equation}
and both pairs of Maxwell's equations can also be presented in the form
\cite{KrLanSus15}%
\begin{equation}
\frac{\partial P_{\mu\nu}}{\partial x^{\lambda}}+\frac{\partial P_{\nu\lambda
}}{\partial x^{\mu}}+\frac{\partial P_{\lambda\mu}}{\partial x^{\nu}}%
=-\frac{4\pi i}{c}e_{\mu\nu\lambda\sigma}j^{\sigma} \label{MaxwellI}%
\end{equation}
in addition to the one given above%
\begin{equation}
\frac{\partial Q^{\mu\nu}}{\partial x^{\nu}}=-\frac{4\pi}{c}j^{\mu}.
\label{MaxwellII}%
\end{equation}
The real part of the first equation traditionally represents the first
(homogeneous) pair of Maxwell's equation and the real part of the second one
gives the remaining pair. In our approach both pairs of Maxwell's equations
appear together (see also \cite{BarutEld}, \cite{Bia:Bia75}, \cite{Bia:Bia13},
\cite{LapUh31}, and \cite{Taylor52} for the case in vacuum). Moreover, a
generalization to complex-valued four-current may naturally represent magnetic
charge and magnetic current not yet observed in nature \cite{Schwinger}.

An important cofactor matrix identity,%
\begin{equation}
P_{\mu\nu}Q^{\nu\lambda}=\left(  \mathbf{F}\cdot\mathbf{G}\right)  \delta
_{\mu}^{\lambda}=\frac{1}{4}\left(  P_{\sigma\tau}Q^{\tau\sigma}\right)
\delta_{\mu}^{\lambda}, \label{CofactorMinkowski}%
\end{equation}
was originally established, in a general form, by Minkowski \cite{MinkowskiI}.
Once again, the dual tensors are given by%
\begin{equation}
P_{\mu\nu}=\left(
\begin{array}
[c]{cc}%
0 & F_{q}\medskip\\
-F_{p} & ie_{pqr}G_{r}%
\end{array}
\right)  ,\quad Q^{\mu\nu}=\left(
\begin{array}
[c]{cc}%
0 & -G_{q}\medskip\\
G_{p} & ie_{pqr}F_{r}%
\end{array}
\right)  , \label{PQblockForm}%
\end{equation}
in block form. A complete list of relevant tensor and matrix identities is
given in appendix~B.

\section{Covariant Derivation of Energy-Momentum Balance Equations}

\subsection{Preliminaries}

As has been announced in \cite{KrLanSus15} (see also \cite{KrLanSus16}), the
covariant form of the differential balance equations can be presented as
follows%
\begin{align}
&  \frac{\partial}{\partial x^{\nu}}\left[  \frac{1}{16\pi}\left(
P_{\mu\lambda}^{\ast}Q^{\lambda\nu}+P_{\mu\lambda}\overset{\ast}{\left.
Q^{\lambda\nu}\right.  }\right)  \right]
\label{CovariantEnergyMomentumBalance}\\
&  \ +\frac{1}{32\pi}\left(  P_{\sigma\tau}^{\ast}\frac{\partial Q^{\tau
\sigma}}{\partial x^{\mu}}+P_{\sigma\tau}\frac{\partial\overset{\ast}{\left.
Q^{\tau\sigma}\right.  }}{\partial x^{\mu}}\right) \nonumber\\
&  \ =-\frac{1}{c}F_{\mu\lambda}j^{\lambda}=\left(
\begin{array}
[c]{c}%
-\mathbf{j}\cdot\mathbf{E}/c\smallskip\\
\rho\mathbf{E}+\mathbf{j}\times\mathbf{B}/c
\end{array}
\right)  .\nonumber
\end{align}
In our complex form, when $\mathbf{F}=\mathbf{E}+i\mathbf{H}$ and
$\mathbf{G}=\mathbf{D}+i\mathbf{B},$ the energy-momentum tensor is given by%
\begin{align}
&  16\pi T_{\mu}{}^{\nu}=P_{\mu\lambda}^{\ast}Q^{\lambda\nu}+P_{\mu\lambda
}\overset{\ast}{\left.  Q^{\lambda\nu}\right.  }%
\label{FourEnergyMomentumTensor}\\
&  =\left(
\begin{array}
[c]{cc}%
\mathbf{F}\cdot\mathbf{G}^{\ast}+\mathbf{F}^{\ast}\cdot\mathbf{G} & 2i\left(
\mathbf{F\times F}^{\ast}\right)  _{q}\bigskip\\
-2i\left(  \mathbf{G\times G}^{\ast}\right)  _{p} & \quad2\left(  F_{p}%
G_{q}^{\ast}+F_{p}^{\ast}G_{q}\right)  -\delta_{pq}\left(  \mathbf{F}%
\cdot\mathbf{G}^{\ast}+\mathbf{F}^{\ast}\cdot\mathbf{G}\right)
\end{array}
\right)  .\nonumber
\end{align}
Here, we point out for the reader's convenience that
\begin{align}
&  i\left(  \mathbf{F\times F}^{\ast}\right)  =2\left(  \mathbf{E\times
H}\right)  ,\qquad i\left(  \mathbf{G\times G}^{\ast}\right)  =2\left(
\mathbf{D\times B}\right)  ,\\
&  \quad\mathbf{F}\cdot\mathbf{G}^{\ast}+\mathbf{F}^{\ast}\cdot\mathbf{G}%
=2\left(  \mathbf{E}\cdot\mathbf{D}+\mathbf{H}\cdot\mathbf{B}\right) \nonumber
\end{align}
and, in real form,%
\begin{equation}
4\pi T_{\mu}{}^{\nu}=\left(
\begin{array}
[c]{cc}%
\left(  \mathbf{E}\cdot\mathbf{D}+\mathbf{H}\cdot\mathbf{B}\right)  /2 &
\left(  \mathbf{E\times H}\right)  _{q}\bigskip\\
-\left(  \mathbf{D\times B}\right)  _{p} & \quad E_{p}D_{q}+H_{p}B_{q}%
-\delta_{pq}\left(  \mathbf{E}\cdot\mathbf{D}+\mathbf{H}\cdot\mathbf{B}%
\right)  /2
\end{array}
\right)  . \label{4EnergyMomentumTensorReal}%
\end{equation}

The covariant form of the differential balance equation allows one to clarify
the meanings of different energy-momentum tensors. For instance, it is worth
noting that the non-symmetric Maxwell and Heaviside form of the $3D$-stress
tensor \cite{Pauli},
\begin{equation}
\widetilde{T}_{pq}=\frac{1}{4\pi}\left(  E_{p}D_{q}+H_{p}B_{q}\right)
-\frac{1}{8\pi}\delta_{pq}\left(  \mathbf{E}\cdot\mathbf{D}+\mathbf{H}%
\cdot\mathbf{B}\right)  , \label{3DEnergyMomentumMaxwellHTensor}%
\end{equation}
appears here in the corresponding \textquotedblleft momentum\textquotedblright%
\ balance equation \cite{Tamm}:%
\begin{align}
&  -\frac{\partial}{\partial t}\left[  \frac{1}{4\pi c}\left(  \mathbf{D}%
\times\mathbf{B}\right)  \right]  _{p}+\frac{\partial\widetilde{T}_{pq}%
}{\partial x_{q}}-\left(  \rho\mathbf{E}+\frac{1}{c}\mathbf{j}\times
\mathbf{B}\right)  _{p}\label{MomentumBalancedRealNonsymmetric}\\
&  \quad+\frac{1}{8\pi}\left(  \mathbf{E}\cdot\frac{\partial\mathbf{D}%
}{\partial x_{p}}-\mathbf{D}\cdot\frac{\partial\mathbf{E}}{\partial x_{p}%
}+\mathbf{H}\cdot\frac{\partial\mathbf{B}}{\partial x_{p}}-\mathbf{B}%
\cdot\frac{\partial\mathbf{H}}{\partial x_{p}}\right)  =0.\nonumber
\end{align}
At the same time, in view of (\ref{MomentumBalancedReal}), use of the form
(\ref{3DEnergyMomentumMaxwellHTensor}) differs from Hertz's symmetric tensors
in (\ref{StressSymComplex}) and (\ref{StressSymReal}) only in the case of
anisotropic media (crystals) \cite{Pauli}, \cite{Tamm39}. Indeed,%
\begin{equation}
8\pi\frac{\partial}{\partial x_{q}}\left(  \widetilde{T}_{pq}-T_{pq}\right)
=\left[  \operatorname{curl}\left(  \mathbf{E\times D}+\mathbf{H}%
\times\mathbf{B}\right)  \right]  _{p}. \label{DivergenceTwoTensors}%
\end{equation}
Moreover, with the help of elementary identities,%
\begin{equation}
\left[  \operatorname{curl}\left(  \mathbf{A\times B}\right)  \right]
_{p}=\frac{\partial}{\partial x_{q}}\left(  A_{p}B_{q}-A_{q}B_{p}\right)
\label{CurlIdentity}%
\end{equation}
and%
\begin{equation}
2\frac{\partial}{\partial x_{q}}\left(  A_{p}B_{q}\right)  =\frac{\partial
}{\partial x_{q}}\left(  A_{p}B_{q}+A_{q}B_{p}\right)  +\left[
\operatorname{curl}\left(  \mathbf{A\times B}\right)  \right]  _{p},
\label{CurlIdentityTwo}%
\end{equation}
one can transform the latter balance equation into its \textquotedblleft
symmetric\textquotedblright\ form, which provides an independent proof of
(\ref{MomentumBalancedReal}).

\subsection{Proof}

The fact that Maxwell's equations can be united with the help of a complex
second rank (anti-symmetric) tensor allows us to utilize the standard
Sturm-Liouville type argument in order to establish the energy-momentum
differential balance equations in covariant form. Indeed, by adding matrix
equation%
\begin{equation}
P_{\mu\lambda}^{\ast}\left(  \frac{\partial Q^{\lambda\nu}}{\partial x^{\nu}%
}=-\frac{4\pi}{c}j^{\lambda}\right)  \label{ProductOne}%
\end{equation}
and its complex conjugate%
\begin{equation}
P_{\mu\lambda}\left(  \frac{\partial\overset{\ast}{\left.  Q^{\lambda\nu
}\right.  }}{\partial x^{\nu}}=-\frac{4\pi}{c}j^{\lambda}\right)
\label{ProductTwo}%
\end{equation}
one gets%
\begin{equation}
P_{\mu\lambda}^{\ast}\frac{\partial Q^{\lambda\nu}}{\partial x^{\nu}}%
+P_{\mu\lambda}\frac{\partial\overset{\ast}{\left.  Q^{\lambda\nu}\right.  }%
}{\partial x^{\nu}}=-\frac{8\pi}{c}F_{\mu\lambda}j^{\lambda}.
\label{CovariantLorentzForce}%
\end{equation}

A simple decomposition,%
\begin{equation}
f\frac{\partial g}{\partial x}=\frac{1}{2}\frac{\partial}{\partial x}\left(
fg\right)  +\frac{1}{2}\left(  f\frac{\partial g}{\partial x}-\frac{\partial
f}{\partial x}g\right)  \label{Decomposition}%
\end{equation}
with $f=P_{\mu\lambda}^{\ast}$ and $g=Q^{\lambda\nu}$ (and their complex
conjugates), results in
\begin{align}
&  \frac{\partial}{\partial x^{\nu}}\left[  \frac{1}{16\pi}\left(
P_{\mu\lambda}^{\ast}Q^{\lambda\nu}+P_{\mu\lambda}\overset{\ast}{\left.
Q^{\lambda\nu}\right.  }\right)  \right] \label{PQForce}\\
&  \ +\frac{1}{16\pi}\left[  \left(  P_{\mu\lambda}^{\ast}\frac{\partial
Q^{\lambda\nu}}{\partial x^{\nu}}-\frac{\partial P_{\mu\lambda}}{\partial
x^{\nu}}\overset{\ast}{\left.  Q^{\lambda\nu}\right.  }\right)  +\left(
\text{c.c.}\right)  \right]  =-\frac{1}{c}F_{\mu\lambda}j^{\lambda}.\nonumber
\end{align}
By a direct substitution, one can verify that
\begin{align}
&  Z_{\mu}=P_{\mu\lambda}^{\ast}\frac{\partial Q^{\lambda\nu}}{\partial
x^{\nu}}-\frac{\partial P_{\mu\lambda}}{\partial x^{\nu}}\overset{\ast
}{\left.  Q^{\lambda\nu}\right.  }=\frac{1}{2}P_{\sigma\tau}^{\ast}%
\frac{\partial Q^{\tau\sigma}}{\partial x^{\mu}}%
\label{CovariantPonderomotiveComplex}\\
&  \quad\ =-\frac{1}{2}\overset{\ast}{\left.  Q^{\sigma\tau}\right.  }%
\frac{\partial P_{\tau\sigma}}{\partial x^{\mu}}=\mathbf{F}^{\ast}\cdot
\frac{\partial\mathbf{G}}{\partial x^{\mu}}-\mathbf{G}^{\ast}\cdot
\frac{\partial\mathbf{F}}{\partial x^{\mu}}.\nonumber
\end{align}
(An independent covariant proof of these identities is given in appendix~C.)
Finally, introducing%
\begin{equation}
16\pi X_{\mu}=Z_{\mu}+Z_{\mu}^{\ast}, \label{CovariantPonderomotiveReal}%
\end{equation}
we obtain (\ref{CovariantEnergyMomentumBalance}) with the explicitly covariant
expression for the ponderomotive force (\ref{PForce}), which completes the proof.

As a result, the covariant energy-momentum balance equation is given by%
\begin{equation}
\frac{\partial}{\partial x^{\nu}}T_{\mu}{}^{\nu}+X_{\mu}=-\frac{1}{c}%
F_{\mu\lambda}j^{\lambda}, \label{CovariantEMBalanceCompact}%
\end{equation}
in a compact form. If these differential balance equations are written for a
stationary medium, then the corresponding equations for moving bodies are
uniquely determined, since the components of a tensor in any inertial
coordinate system can be derived by a proper Lorentz transformation
\cite{Pauli}.

\section{Covariant Derivation of Angular Momentum Balance}

By definition, $x_{\mu}=g_{\mu\nu}x^{\nu}=\left(  ct,-\mathbf{r}\right)  $ and
$T_{\mu\lambda}=T_{\mu}{}^{\nu}g_{\nu\lambda},$ where $g_{\mu\nu}%
=$diag$\left(  1,-1,-1,-1\right)  =\partial x_{\mu}/\partial x^{\nu}.$ In view
of (\ref{CovariantEMBalanceCompact}), we derive%
\begin{align}
&  \frac{\partial}{\partial x^{\nu}}\left(  x_{\lambda}T_{\mu}{}^{\nu}-x_{\mu
}T_{\lambda}{}^{\nu}\right)  =\left(  T_{\mu\lambda}-T_{\lambda\mu}\right)
\label{CovariantAngularMomentum}\\
&  -\left(  x_{\lambda}X_{\mu}-x_{\mu}X_{\lambda}\right)  -\frac{1}{c}\left(
x_{\lambda}F_{\mu\nu}-x_{\mu}F_{\lambda\nu}\right)  j^{\nu}\nonumber
\end{align}
as a required differential balance equation.

With the help of familiar dual relations (\ref{B04}), one can get another
covariant form of the angular momentum balance equation:%
\begin{align}
&  \frac{\partial}{\partial x^{\nu}}\left(  e^{\mu\lambda\sigma\tau}x_{\sigma
}T_{\tau}{}^{\nu}\right)  +e^{\mu\lambda\sigma\tau}T_{\sigma\tau
}\label{CovariantMomentumBalance}\\
&  \ +e^{\mu\lambda\sigma\tau}x_{\sigma}X_{\tau}{}+\frac{1}{c}e^{\mu
\lambda\sigma\tau}x_{\sigma}F_{\tau\nu}j^{\nu}=0^{\mu\lambda}.\nonumber
\end{align}
In $3D$-form, the latter relation can be reduced to
(\ref{VectorMomentumBalance})--(\ref{TensorMomentum}).

Indeed, when $\mu=0$ and $\lambda=p=1,2,3,$ one gets%
\begin{align}
&  -\frac{1}{4\pi c}\frac{\partial}{\partial t}\left[  e_{pqr}x_{q}\left(
\mathbf{D}\times\mathbf{B}\right)  _{r}\right]  +\frac{\partial}{\partial
x_{s}}\left(  e_{pqr}x_{q}\widetilde{T}_{rs}\right) \label{4Dto3DAngMom}\\
&  \qquad+e_{pqr}\widetilde{T}_{qr}+e_{pqr}x_{q}\left(  X_{r}+Y_{r}\right)
=0,\nonumber
\end{align}
where $-\mathbf{Y}=\rho\mathbf{E}+\mathbf{j}\times\mathbf{B}/c$ is the
familiar Lorentz force. Substitution, $\widetilde{T}_{rs}=T_{rs}+\left(
\widetilde{T}_{rs}-T_{rs}\right)  ,$ results in (\ref{VectorMomentumBalance})
in view of identity (\ref{DivergenceTwoTensors}). The remaining cases, when
$\mu,\nu=p,q=1,2,3,$ can be analyzed in a similar fashion. In $3D$-form, the
corresponding equations can be reduced to (\ref{EnergyBalance}) and
(\ref{MomentumBalancedRealNonsymmetric}). Details are left to the reader.

Thus the angular momentum law has the form of a local balance equation, not a
conservation law, since in general the energy-momentum tensor will not be
symmetric \cite{deGrootSuttorp}. A torque, for instance, may occur, which
cannot be compensated for by a change in the electromagnetic angular momentum,
though not in contradiction with experiment \cite{Pauli}.

\section{Transformation Laws of Complex Electromagnetic Fields}

Let $\mathbf{v}$ be a constant real-valued velocity vector representing
uniform motion of one frame of reference with respect to another one. Let us
consider the following orthogonal decompositions,%
\begin{equation}
\mathbf{F}=\mathbf{F}_{\parallel}+\mathbf{F}_{\perp},\qquad\mathbf{G}%
=\mathbf{G}_{\parallel}+\mathbf{G}_{\perp}, \label{FGdecompositions}%
\end{equation}
such that our complex vectors $\left\{  \mathbf{F}_{\parallel},\mathbf{G}%
_{\parallel}\right\}  $ are collinear with the velocity vector $\mathbf{v}$
and $\left\{  \mathbf{F}_{\perp},\mathbf{G}_{\perp}\right\}  $ are
perpendicular to it (Figure~1). The Lorentz transformation of electric and
magnetic fields $\left\{  \mathbf{E,D},\mathbf{H},\mathbf{B}\right\}  $ take
the following complex form%
\begin{equation}
\mathbf{F}_{\parallel}^{\prime}=\mathbf{F}_{\parallel},\qquad\mathbf{G}%
_{\parallel}^{\prime}=\mathbf{G}_{\parallel} \label{ParallelTransform}%
\end{equation}
and%
\begin{equation}
\mathbf{F}_{\perp}^{\prime}=\frac{\mathbf{F}_{\perp}-\dfrac{i}{c}\left(
\mathbf{v\times G}\right)  }{\sqrt{1-v^{2}/c^{2}}},\qquad\mathbf{G}_{\perp
}^{\prime}=\frac{\mathbf{G}_{\perp}-\dfrac{i}{c}\left(  \mathbf{v\times
F}\right)  }{\sqrt{1-v^{2}/c^{2}}}. \label{PerpendicularTransform}%
\end{equation}
Although this transformation was found by Lorentz, it was Minkowski who
realized that this is the law of transformation of the second rank
anti-symmetric four-tensors \cite{LorentzEinsteinMinkowski}, \cite{MinkowskiI}%
; a brief historical overview is given in \cite{Pauli}.) This complex
$3D$-form of the Lorentz transformation of electric and magnetic fields was
known to Minkowski (1908), but apparently only in vacuum, when $\mathbf{G}%
=\mathbf{F}$ (see also \cite{TerRyb}). Moreover,%
\begin{equation}
\mathbf{r}_{\parallel}^{\prime}=\frac{\mathbf{r}_{\parallel}-\mathbf{v}%
t}{\sqrt{1-v^{2}/c^{2}}},\qquad\mathbf{r}_{\perp}^{\prime}=\mathbf{r}_{\perp
},\qquad t^{\prime}=\frac{t-\left(  \mathbf{v}\cdot\mathbf{r}\right)  /c^{2}%
}{\sqrt{1-v^{2}/c^{2}}}, \label{LorentzCoordinateTransform}%
\end{equation}
in the same notation \cite{Pauli}.
\begin{figure}[htbp]
\centering%
\scalebox{.75}%
{\includegraphics{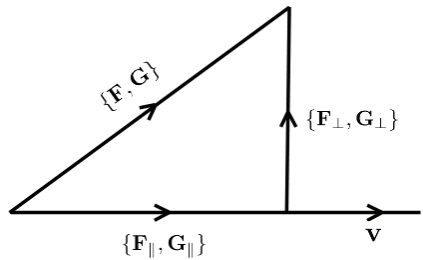}}
\caption{Complex electromagnetic fields decomposition.}
\end{figure}

The latter equations can be rewritten as follows%
\begin{equation}
\mathbf{r}^{\prime}=\mathbf{r}+\left[  \left(  \gamma-1\right)  \frac
{\mathbf{v}\cdot\mathbf{r}}{v^{2}}-\gamma t\right]  \mathbf{v},\qquad
t^{\prime}=\gamma\left(  t-\frac{\mathbf{v}\cdot\mathbf{r}}{c^{2}}\right)  ,
\label{LorentzCoordinateTransformShort}%
\end{equation}
where $\gamma=\left(  1-v^{2}/c^{2}\right)  ^{-1/2}.$ In a similar fashion,
one gets%
\begin{equation}
\mathbf{F}^{\prime}=\gamma\left(  \mathbf{F}-\dfrac{i}{c}\mathbf{v\times
G}\right)  -\left(  \gamma-1\right)  \frac{\mathbf{v}\cdot\mathbf{F}}{v^{2}%
}\mathbf{v}, \label{LorentzFTransform}%
\end{equation}%
\begin{equation}
\mathbf{G}^{\prime}=\gamma\left(  \mathbf{G}-\dfrac{i}{c}\mathbf{v\times
F}\right)  -\left(  \gamma-1\right)  \frac{\mathbf{v}\cdot\mathbf{G}}{v^{2}%
}\mathbf{v}, \label{LorentzGTransform}%
\end{equation}
as a compact $3D$-version of the Lorentz transformation for the complex
electromagnetic fields.

In complex four-tensor form,%
\begin{equation}
\left.  Q^{\prime}\right.  ^{\mu\nu}\left(  x^{\prime}\right)  =\Lambda
_{\ \sigma}^{\mu}\Lambda_{\ \tau}^{\nu}Q^{\sigma\tau}\left(  x\right)  ,\qquad
x^{\prime}=\Lambda x. \label{ComplexFieldTransform}%
\end{equation}
Although Minkowski considered the transformation of electric and magnetic
fields in a complex $3D$-vector form, see Eqs.~(8)--(9) and (15) in
\cite{MinkowskiI} (or Eqs.~(25.5)--(25.6) in \cite{LanLif2}), he seems never
to have combined the corresponding four-tensors into the complex
forms~(\ref{ComplexQ})--(\ref{ComplexP}). In the second article
\cite{MinkowskiII}, Max Born, who used Minkowski's notes, didn't mention the
complex fields. As a result, the complex field tensor seems only to have
appeared, for the first time, in \cite{LapUh31} (see also \cite{Taylor52}).
The complex identity, $\mathbf{F}\cdot\mathbf{G}=~$invariant$~$under a
similarity transformation, follows from Minkowski's determinant relations
(\ref{B18})--(\ref{B20}).
\begin{figure}[htbp]
\centering%
\scalebox{.75}%
{\includegraphics{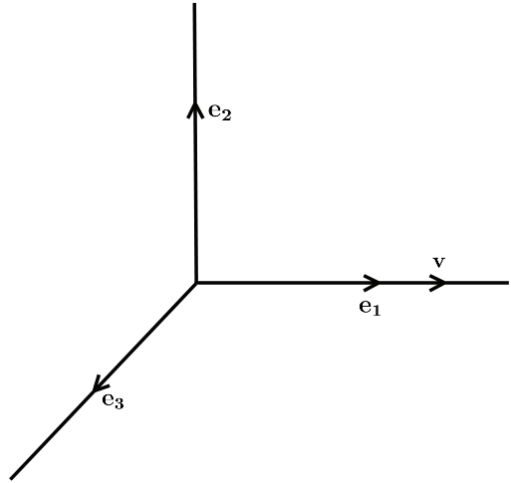}}
\caption{Example of moving frame velocity.}
\end{figure}

\noindent{\textbf{Example}}. Let $\left\{  \mathbf{e}_{k}\right\}  _{k=1}^{3}$
be an orthonormal basis in $\left.
\mathbb{R}
\right.  ^{3}.$ We choose $\mathbf{v}=v\mathbf{e}_{1}$ and write $x^{\prime
\mu}=\Lambda_{\ \nu}^{\mu}x^{\nu}$ with%
\begin{equation}
\Lambda_{\ \nu}^{\mu}=\left(
\begin{array}
[c]{cccc}%
\gamma & -\beta\gamma & 0 & 0\\
-\beta\gamma & \gamma & 0 & 0\\
0 & 0 & 1 & 0\\
0 & 0 & 0 & 1
\end{array}
\right)  ,\qquad\quad\beta=\frac{v}{c},\quad\gamma=\frac{1}{\sqrt{1-\beta^{2}%
}} \label{LorentzBoost}%
\end{equation}
for the corresponding Lorentz boost (Figure~2). In view of
(\ref{ComplexFieldTransform}), by matrix multiplication one gets%
\begin{align}
&  \left(
\begin{array}
[c]{cccc}%
\gamma & -\beta\gamma & 0 & 0\\
-\beta\gamma & \gamma & 0 & 0\\
0 & 0 & 1 & 0\\
0 & 0 & 0 & 1
\end{array}
\right)  \left(
\begin{array}
[c]{cccc}%
0 & -G_{1} & -G_{2} & -G_{3}\\
G_{1} & 0 & iF_{3} & -iF_{2}\\
G_{2} & -iF_{3} & 0 & iF_{1}\\
G_{3} & iF_{2} & -iF_{1} & 0
\end{array}
\right)  \left(
\begin{array}
[c]{cccc}%
\gamma & -\beta\gamma & 0 & 0\\
-\beta\gamma & \gamma & 0 & 0\\
0 & 0 & 1 & 0\\
0 & 0 & 0 & 1
\end{array}
\right) \nonumber\\
&  \qquad=\left(
\begin{array}
[c]{cccc}%
0 & -G_{1} & -\gamma G_{2}-i\beta\gamma F_{3} & -\gamma G_{3}+i\beta\gamma
F_{2}\\
G_{1} & 0 & \beta\gamma G_{2}+i\gamma F_{3} & \beta\gamma G_{3}-i\gamma
F_{2}\\
\gamma G_{2}+i\beta\gamma F_{3} & -\beta\gamma G_{2}-i\gamma F_{3} & 0 &
iF_{1}\\
\gamma G_{3}-i\beta\gamma F_{2} & -\beta\gamma G_{3}+i\gamma F_{2} & -iF_{1} &
0
\end{array}
\right)  . \label{MartixMuttiplication}%
\end{align}
Thus $G_{1}^{\prime}=G_{1}$ and%
\begin{align}
G_{2}^{\prime}  &  =\gamma G_{2}+i\beta\gamma F_{3}=\frac{G_{2}+i\left(
v/c\right)  F_{3}}{\sqrt{1-v^{2}/c^{2}}}=\frac{G_{2}-\dfrac{i}{c}\left(
\mathbf{v\times F}\right)  _{2}}{\sqrt{1-v^{2}/c^{2}}},\label{G2G3}\\
G_{3}^{\prime}  &  =\gamma G_{3}-i\beta\gamma F_{2}=\frac{G_{3}-i\left(
v/c\right)  F_{2}}{\sqrt{1-v^{2}/c^{2}}}=\frac{G_{3}-\dfrac{i}{c}\left(
\mathbf{v\times F}\right)  _{3}}{\sqrt{1-v^{2}/c^{2}}}.\nonumber
\end{align}
In a similar fashion, $F_{1}^{\prime}=F_{1}$ and%
\begin{align}
F_{2}^{\prime}  &  =\gamma F_{2}+i\beta\gamma G_{3}=\frac{F_{2}-\dfrac{i}%
{c}\left(  \mathbf{v\times G}\right)  _{2}}{\sqrt{1-v^{2}/c^{2}}}%
,\label{F2F3}\\
F_{3}^{\prime}  &  =\gamma F_{3}-i\beta\gamma G_{2}=\frac{F_{3}-\dfrac{i}%
{c}\left(  \mathbf{v\times G}\right)  _{3}}{\sqrt{1-v^{2}/c^{2}}}.\nonumber
\end{align}
The latter relations are in agreement with the field transformations
(\ref{ParallelTransform})--(\ref{PerpendicularTransform}).

In block form, one gets%
\begin{equation}
\left(
\begin{array}
[c]{c}%
F_{1}^{\prime}\\
F_{2}^{\prime}\\
G_{3}^{\prime}\\
G_{2}^{\prime}\\
F_{3}^{\prime}\\
G_{1}^{\prime}%
\end{array}
\right)  =\left(
\begin{array}
[c]{cccccc}%
1 & 0 & 0 & 0 & 0 & 0\\
0 & \cos\left(  i\theta\right)  & \sin\left(  i\theta\right)  & 0 & 0 & 0\\
0 & -\sin\left(  i\theta\right)  & \cos\left(  i\theta\right)  & 0 & 0 & 0\\
0 & 0 & 0 & \cos\left(  i\theta\right)  & \sin\left(  i\theta\right)  & 0\\
0 & 0 & 0 & -\sin\left(  i\theta\right)  & \cos\left(  i\theta\right)  & 0\\
0 & 0 & 0 & 0 & 0 & 1
\end{array}
\right)  \left(
\begin{array}
[c]{c}%
F_{1}\\
F_{2}\\
G_{3}\\
G_{2}\\
F_{3}\\
G_{1}%
\end{array}
\right)  , \label{EMfieldsTransformMatrix}%
\end{equation}
where, by definition,%
\begin{equation}
\cos\left(  i\theta\right)  =\gamma=\frac{1}{\sqrt{1-\beta^{2}}},\qquad
\sin\left(  i\theta\right)  =i\beta\gamma=\frac{i\beta}{\sqrt{1-\beta^{2}}%
},\qquad\beta=\frac{v}{c}.
\end{equation}
As a result, the transformation law of the complex electromagnetic fields
$\left\{  \mathbf{F},\mathbf{G}\right\}  $ under the Lorentz boost can be
thought of as a complex rotation in $\left.
\mathbb{C}
\right.  ^{6},$ corresponding to a reducible representation of the
one-parameter subgroup of $SO\left(  3,%
\mathbb{C}
\right)  .$ (Cyclic permutation of the spatial indices cover the two remaining
cases; see also \cite{TerRyb}.)

\section{Material Equations, Potentials, and Energy-Momentum Tensor for Moving
Isotropic Media}

Electromagnetic phenomena in moving media are important in relativistic
astrophysics, the study of accelerated plasma clusters and high-energy
electron beams \cite{BolStolUFN75}, \cite{BolStolSb83}, \cite{FleishTop13},
\cite{ToptyginII}.

\subsection{Material equations}

Minkowski's field- and connecting-equations \cite{MinkowskiI},
\cite{MinkowskiII} were derived from the corresponding laws for the bodies at
rest by means of a Lorentz transformation (see \cite{BolStolUFN75},
\cite{Denisov89}, \cite{deGrootSuttorp}, \cite{LanLif8},
\cite{Nesterenko2JMP16}, \cite{Pauli}, \cite{ToptyginII}). Explicitly
covariant forms, which are applicable both in the rest frame and for moving
media, are analyzed in \cite{BolStolUFN75}, \cite{BolStolSb83},
\cite{deGrootSuttorp}, \cite{JauchWatson48a}, \cite{JauchWatson48b},
\cite{Nesterenko2JMP16}, \cite{Papas}, \cite{Pauli}, \cite{Riazanov57},
\cite{Ryazanov58}, \cite{TerRyb}, \cite{ToptyginII} (see also the references
therein). In standard notation,%
\begin{equation}
\beta=v/c,\qquad\gamma=\left(  1-\beta^{2}\right)  ^{-1/2},\qquad v=\left\vert
\mathbf{v}\right\vert ,\qquad\kappa=\varepsilon\mu-1, \label{BetaGamma}%
\end{equation}
one can write \cite{BolStolUFN75}, \cite{BolStolSb83}, \cite{Denisov89},
\cite{ToptyginII}:
\begin{align}
\mathbf{D}  &  =\varepsilon\mathbf{E}+\frac{\kappa\gamma^{2}}{\mu}\left[
\beta^{2}\mathbf{E}-\dfrac{\mathbf{v}}{c^{2}}\left(  \mathbf{v\cdot E}\right)
+\dfrac{1}{c}\left(  \mathbf{v\times B}\right)  \right]
,\label{3DMaterialEqs}\\
\mathbf{H}  &  =\frac{1}{\mu}\mathbf{B}+\frac{\kappa\gamma^{2}}{\mu}\left[
-\beta^{2}\mathbf{B}+\dfrac{\mathbf{v}}{c^{2}}\left(  \mathbf{v\cdot
B}\right)  +\dfrac{1}{c}\left(  \mathbf{v\times E}\right)  \right]  .\nonumber
\end{align}
In covariant form, these relations are given by%
\begin{align}
R^{\lambda\nu}  &  =\epsilon^{\lambda\nu\sigma\tau}F_{\sigma\tau}=\frac{1}%
{2}\left(  \epsilon^{\lambda\nu\sigma\tau}-\epsilon^{\lambda\nu\tau\sigma
}\right)  F_{\sigma\tau}\label{CovariantMaterialEqs}\\
&  =\frac{1}{4}\left(  \epsilon^{\lambda\nu\sigma\tau}-\epsilon^{\lambda
\nu\tau\sigma}+\epsilon^{\nu\lambda\tau\sigma}-\epsilon^{\nu\lambda\sigma\tau
}\right)  F_{\sigma\tau}\nonumber
\end{align}
(see \cite{BolRuk60}, \cite{BolStolUFN75}, \cite{BolStolSb83},
\cite{JauchWatson48a}, \cite{JauchWatson48b}, \cite{Riazanov57},
\cite{Ryazanov58}, \cite{ToptyginII} and the references therein). Here,%
\begin{equation}
\epsilon^{\lambda\nu\sigma\tau}=\frac{1}{\mu}\left(  g^{\lambda\sigma}+\kappa
u^{\lambda}u^{\sigma}\right)  \left(  g^{\nu\tau}+\kappa u^{\nu}u^{\tau
}\right)  =\epsilon^{\nu\lambda\tau\sigma} \label{FourPermeabilityTensor}%
\end{equation}
is the four-tensor of electric and magnetic permeabilities\footnote{Originally
introduced by Tamm for a general case of the moving anisotropic medium
\cite{Tamm24}, \cite{Tamm25}.} and%
\begin{equation}
u^{\lambda}=\left(  \gamma,\gamma\mathbf{v}/c\right)  ,\qquad u^{\lambda
}u_{\lambda}=1 \label{FourVelocity}%
\end{equation}
is the four-velocity of the medium (\cite{Riazanov57}, \cite{Ryazanov58}, a
computer algebra verification of these relations is given in
\cite{LanfearMath}). In a complex covariant form,%
\begin{equation}
\left(  Q^{\mu\nu}+\overset{\ast}{\left.  Q^{\mu\nu}\right.  }\right)
=\epsilon^{\mu\nu\sigma\tau}\left(  P_{\sigma\tau}+\overset{\ast}{\left.
P_{\sigma\tau}\right.  }\right)  . \label{CovariantMaterialEqsComplex}%
\end{equation}
In view of (\ref{CovariantMaterialEqs}) and (\ref{B1})--(\ref{B2}), we get%
\begin{equation}
Q^{\mu\nu}=\left(  \epsilon^{\mu\nu\sigma\tau}-\frac{i}{2}e^{\mu\nu\sigma\tau
}\right)  F_{\sigma\tau},\qquad P_{\mu\nu}=\left(  \delta_{\mu}^{\lambda
}\delta_{\nu}^{\rho}-\frac{i}{2}e_{\mu\nu\sigma\tau}\epsilon^{\sigma
\tau\lambda\rho}\right)  F_{\lambda\rho}, \label{QPmaterialF}%
\end{equation}
in terms of the real-valued electromagnetic field tensor.

\subsection{Potentials}

In practice, one can choose%
\begin{equation}
F_{\sigma\tau}=\frac{\partial A_{\tau}}{\partial x^{\sigma}}-\frac{\partial
A_{\sigma}}{\partial x^{\tau}}, \label{4VectorPotential}%
\end{equation}
for the real-valued four-vector potential $A_{\lambda}\left(  x\right)  .$
Then%
\begin{align*}
\partial_{\nu}Q^{\lambda\nu}  &  =\epsilon^{\lambda\nu\sigma\tau}\partial
_{\nu}\left(  \partial_{\sigma}A_{\tau}-\partial_{\tau}A_{\sigma}\right)
-\frac{i}{2}e^{\lambda\nu\sigma\tau}\partial_{\nu}\left(  \partial_{\sigma
}A_{\tau}-\partial_{\tau}A_{\sigma}\right) \\
&  =\frac{1}{\mu}\left(  g^{\lambda\sigma}+\kappa u^{\lambda}u^{\sigma
}\right)  \left(  g^{\nu\tau}+\kappa u^{\nu}u^{\tau}\right)  \partial_{\nu
}\left(  \partial_{\sigma}A_{\tau}-\partial_{\tau}A_{\sigma}\right)
\end{align*}
by (\ref{FourPermeabilityTensor}). Substitution into Maxwell's equations
(\ref{MaxwellII}) or (\ref{MaxwellI}) results in
\begin{align}
&  \left(  g^{\lambda\sigma}+\kappa u^{\lambda}u^{\sigma}\right)  \left\{
-\left[  \partial^{\tau}\partial_{\tau}+\kappa\left(  u^{\tau}\partial_{\tau
}\right)  ^{2}\right]  A_{\sigma}\right. \label{CovariantAu}\\
&  \qquad\left.  +\partial_{\sigma}\left(  \partial^{\tau}A_{\tau}+\kappa
u^{\nu}u^{\tau}\partial_{\nu}A_{\tau}\right)  ^{\ }\!\right\}  =-\frac{4\pi
\mu}{c}j^{\lambda},\nonumber
\end{align}
where $-\partial^{\tau}\partial_{\tau}=-g^{\sigma\tau}\partial_{\sigma
}\partial_{\tau}=\Delta-\left(  \partial/c\partial t\right)  ^{2}$ is the
D'Alembert operator. In view of an inverse matrix identity,%
\begin{equation}
\left(  g_{\lambda\rho}-\frac{\kappa}{1+\kappa}u_{\lambda}u_{\rho}\right)
\left(  g^{\lambda\sigma}+\kappa u^{\lambda}u^{\sigma}\right)  =\delta_{\rho
}^{\sigma}, \label{InverseA}%
\end{equation}
the latter equations take the form\footnote{Equations (\ref{4VectorPotential})
and (\ref{CovariantAuj}), together with the gauge condition
(\ref{CovariantAuGauge}), may be considered as the fundamentals of the theory
\cite{JauchWatson48a}. Our complex fields are given by (\ref{QPmaterialF}).}%
\begin{align}
&  \left[  \partial^{\tau}\partial_{\tau}+\kappa\left(  u^{\tau}\partial
_{\tau}\right)  ^{2}\right]  A_{\sigma}-\partial_{\sigma}\left(
\partial^{\tau}A_{\tau}+\kappa u^{\nu}u^{\tau}\partial_{\nu}A_{\tau}\right)
\label{CovariantAuj}\\
&  \qquad=\frac{4\pi\mu}{c}\left(  g_{\sigma\lambda}-\frac{\kappa}{1+\kappa
}u_{\sigma}u_{\lambda}\right)  j^{\lambda}.\nonumber
\end{align}
Subject to the subsidiary condition,%
\begin{equation}
\partial^{\tau}A_{\tau}+\kappa u^{\nu}u^{\tau}\partial_{\nu}A_{\tau}=\left(
g^{\nu\tau}+\kappa u^{\nu}u^{\tau}\right)  \partial_{\nu}A_{\tau}=0,
\label{CovariantAuGauge}%
\end{equation}
these equations were studied in detail for the sake of development of the
phenomenological classical and quantum electrodynamics in a moving medium (see
\cite{Bol09}, \cite{BolRuk60}, \cite{BolStolUFN75}, \cite{BolStolSb83},
\cite{Papas}, \cite{Riazanov57}, \cite{Riazanov58}, \cite{Ryazanov58},
\cite{ToptyginII} and the references therein). In particular, Green's function
of the photon in a moving medium was studied in \cite{JauchWatson48a},
\cite{Riazanov57}, \cite{Riazanov58} (with applications to quantum electrodynamics).

\subsection{Hertz's tensor and vectors}

We follow \cite{BolStolUFN75}, \cite{BolStolSb83}, \cite{ToptyginII} with
somewhat different details. The substitution,%
\begin{equation}
A^{\mu}\left(  x\right)  =\left(  \frac{\kappa}{1+\kappa}u^{\mu}u_{\lambda
}-\delta_{\lambda}^{\mu}\right)  \partial_{\sigma}Z^{\lambda\sigma}\left(
x\right)  \label{AZHertz}%
\end{equation}
(a generalization of Hertz's potentials for a moving medium
\cite{BolStolUFN75}, \cite{ToptyginII}), into the gauge condition
(\ref{CovariantAuGauge}) results in $Z^{\lambda\sigma}=-Z^{\sigma\lambda},$ in
view of%
\begin{align*}
&  \left(  g_{\nu\mu}+\kappa u_{\nu}u_{\mu}\right)  \partial^{\nu}A^{\mu}\\
&  \quad=\left(  g_{\nu\mu}+\kappa u_{\nu}u_{\mu}\right)  \left(  \frac
{\kappa}{1+\kappa}u^{\mu}u_{\lambda}-\delta_{\lambda}^{\mu}\right)
\partial^{\nu}\partial_{\sigma}Z^{\lambda\sigma}\\
&  \qquad=-g_{\nu\lambda}\partial^{\nu}\partial_{\sigma}Z^{\lambda\sigma
}=-\partial_{\lambda}\partial_{\sigma}Z^{\lambda\sigma}\equiv0.
\end{align*}
Then, equations (\ref{CovariantAuj}) take the form%
\begin{equation}
\left[  \partial^{\tau}\partial_{\tau}+\kappa\left(  u^{\tau}\partial_{\tau
}\right)  ^{2}\right]  \partial_{\sigma}Z^{\lambda\sigma}=-\frac{4\pi\mu}%
{c}j^{\lambda}. \label{dZjHertzEquation}%
\end{equation}
Indeed, the left-hand side of (\ref{CovariantAuj}) is given by%
\begin{align*}
&  \left[  \partial^{\tau}\partial_{\tau}+\kappa\left(  u^{\tau}\partial
_{\tau}\right)  ^{2}\right]  A_{\sigma}=\left[  \partial^{\tau}\partial_{\tau
}+\kappa\left(  u^{\tau}\partial_{\tau}\right)  ^{2}\right]  g_{\sigma\mu
}A^{\mu}\\
&  \quad=\left[  \partial^{\tau}\partial_{\tau}+\kappa\left(  u^{\tau}%
\partial_{\tau}\right)  ^{2}\right]  g_{\sigma\mu}\left(  \frac{\kappa
}{1+\kappa}u^{\mu}u_{\lambda}-\delta_{\lambda}^{\mu}\right)  \partial_{\rho
}Z^{\lambda\rho}\\
&  \quad=\left[  \partial^{\tau}\partial_{\tau}+\kappa\left(  u^{\tau}%
\partial_{\tau}\right)  ^{2}\right]  \left(  \frac{\kappa}{1+\kappa}u_{\sigma
}u_{\lambda}-g_{\sigma\lambda}\right)  \partial_{\rho}Z^{\lambda\rho}\\
&  \qquad=\frac{4\pi\mu}{c}\left(  g_{\sigma\lambda}-\frac{\kappa}{1+\kappa
}u_{\sigma}u_{\lambda}\right)  j^{\lambda},
\end{align*}
from which the result follows due to (\ref{InverseA}).

Finally, with the help of the standard substitution,%
\begin{equation}
j^{\lambda}=c\partial_{\sigma}p^{\lambda\sigma},\qquad p^{\lambda\sigma
}=-p^{\sigma\lambda} \label{pHertz}%
\end{equation}
(in view of $\partial_{\lambda}j^{\lambda}=c\partial_{\lambda}\partial
_{\sigma}p^{\lambda\sigma}\equiv0),$ we arrive at%
\begin{equation}
\partial_{\sigma}\left\{  \left[  \partial^{\tau}\partial_{\tau}+\kappa\left(
u^{\tau}\partial_{\tau}\right)  ^{2}\right]  Z^{\lambda\sigma}+4\pi\mu
p^{\lambda\sigma}\right\}  =0. \label{DZpHertz}%
\end{equation}
Therefore, one can choose%
\begin{equation}
\left[  \partial^{\tau}\partial_{\tau}+\kappa\left(  u^{\tau}\partial_{\tau
}\right)  ^{2}\right]  Z^{\lambda\nu}=-4\pi p^{\lambda\nu}. \label{ZpHertz}%
\end{equation}
Here, by definition,%
\begin{equation}
p^{\lambda\nu}=\left(
\begin{array}
[c]{cccc}%
0 & -p_{1} & -p_{2} & -p_{3}\\
p_{1} & 0 & m_{3} & -m_{2}\\
p_{2} & -m_{3} & 0 & m_{1}\\
p_{3} & m_{2} & -m_{1} & 0
\end{array}
\right)  \label{pTensor}%
\end{equation}
is an anti-symmetric four-tensor \cite{BolStolUFN75}, \cite{BolStolSb83},
\cite{ToptyginII}. The \textquotedblleft electric\textquotedblright\ and
\textquotedblleft magnetic\textquotedblright\ Hertz vectors, $\mathbf{Z}%
^{\left(  e\right)  }$ and $\mathbf{Z}^{\left(  m\right)  },$ are also
introduced in terms of a single four-tensor,%
\begin{equation}
Z^{\lambda\nu}=\left(
\begin{array}
[c]{cccc}%
0 & Z_{1}^{\left(  e\right)  } & Z_{2}^{\left(  e\right)  } & Z_{3}^{\left(
e\right)  }\\
-Z_{1}^{\left(  e\right)  } & 0 & -Z_{3}^{\left(  m\right)  } & Z_{2}^{\left(
m\right)  }\\
-Z_{2}^{\left(  e\right)  } & Z_{3}^{\left(  m\right)  } & 0 & -Z_{1}^{\left(
m\right)  }\\
-Z_{3}^{\left(  e\right)  } & -Z_{2}^{\left(  m\right)  } & Z_{1}^{\left(
m\right)  } & 0
\end{array}
\right)  . \label{ZTensor}%
\end{equation}
In view of (\ref{AZHertz}), for the four-vector potential, $A^{\lambda
}=\left(  \varphi,\mathbf{A}\right)  ,$ we obtain%
\begin{equation}
\varphi=-\left(  1-\frac{\kappa\gamma^{2}}{1+\kappa}\right)
\operatorname{div}\mathbf{Z}^{\left(  e\right)  }+\frac{\kappa\gamma^{2}%
}{\left(  1+\kappa\right)  c}\mathbf{v}\cdot\left(  \frac{\partial
\mathbf{Z}^{\left(  e\right)  }}{c\partial t}+\operatorname{curl}%
\mathbf{Z}^{\left(  m\right)  }\right)  \label{phiZ}%
\end{equation}
and%
\begin{align}
&  \mathbf{A}=\frac{\partial\mathbf{Z}^{\left(  e\right)  }}{c\partial
t}+\operatorname{curl}\mathbf{Z}^{\left(  m\right)  }\label{AZ}\\
&  \quad+\frac{\kappa\gamma^{2}\mathbf{v}}{\left(  1+\kappa\right)  c^{2}%
}\left[  c\operatorname{div}\mathbf{Z}^{\left(  e\right)  }+\frac{\partial
}{c\partial t}\left(  \mathbf{v}\cdot\mathbf{Z}^{\left(  e\right)  }\right)
+\mathbf{v}\cdot\operatorname{curl}\mathbf{Z}^{\left(  m\right)  }\right]
.\nonumber
\end{align}
Then, equations (\ref{ZpHertz}) take the form%
\begin{equation}
\left[  \partial^{\tau}\partial_{\tau}+\kappa\left(  u^{\tau}\partial_{\tau
}\right)  ^{2}\right]  \mathbf{Z}^{\left(  e\right)  }=4\pi\mu\mathbf{p}%
,\qquad\left[  \partial^{\tau}\partial_{\tau}+\kappa\left(  u^{\tau}%
\partial_{\tau}\right)  ^{2}\right]  \mathbf{Z}^{\left(  m\right)  }=4\pi
\mu\mathbf{m} \label{ZEMpm}%
\end{equation}
and, for the four-current, $j^{\lambda}=\left(  c\rho,\mathbf{j}\right)  ,$
one gets%
\begin{equation}
\rho=-\operatorname{div}\mathbf{p},\qquad\mathbf{j}=\frac{\partial\mathbf{p}%
}{\partial t}+c\operatorname{curl}\mathbf{m} \label{rhojpm}%
\end{equation}
(see \cite{BolStolUFN75}, \cite{BolStolSb83}, \cite{ToptyginII} for more details).

The Hertz vector and tensor potentials, for a moving medium and at rest, were
utilized in \cite{BolStolUFN75}, \cite{BolStolSb83}, \cite{Fock65},
\cite{Kann87}, \cite{Tamm}, \cite{ToptyginII}, \cite{VinRudSuxBook79} (see
also the references therein). Many classical problems of radiation and
propagation can be consistently solved by using these potentials.

\subsection{Energy-momentum tensor}

In the case of the covariant version of the energy-momentum tensor given by
(\ref{FourEnergyMomentumTensor}), the differential balance equations under
consideration are independent of the particular choice of the frame of
reference. Therefore, our relations (\ref{QPmaterialF}) are useful for
derivation of the expressions for the energy-momentum tensor and the
ponderomotive force for moving bodies from those for bodies at rest which were
extensively studied in the literature. For example, one gets%
\begin{equation}
4\pi T_{\mu}{}^{\nu}=F_{\mu\lambda}\epsilon^{\lambda\nu\sigma\tau}%
F_{\sigma\tau}+\frac{1}{4}\delta_{\mu}^{\nu}F_{\sigma\tau}\epsilon^{\sigma
\tau\lambda\rho}F_{\lambda\rho} \label{CovariantEMTensorMoving}%
\end{equation}
with the help of (\ref{CovariantMaterialEqs})--(\ref{FourPermeabilityTensor})
and (\ref{B10}) (see also \cite{Tamm25}).

\section{Real versus Complex Lagrangians}

In modern presentations of the classical and quantum field theories, the
Lagrangian approach is usually utilized.

\subsection{Complex forms}

We introduce two quadratic \textquotedblleft Lagrangian\textquotedblright%
\ densities%
\begin{align}
\mathcal{L}_{0}  &  =\mathcal{L}_{0}^{\ast}=\frac{1}{2}\left(  P_{\sigma\tau
}Q^{\tau\sigma}+P_{\sigma\tau}^{\ast}\overset{\ast}{\left.  Q^{\tau\sigma
}\right.  }\right) \label{Lagrangian0}\\
&  =\frac{i}{4}e^{\sigma\tau\kappa\rho}\left(  P_{\sigma\tau}P_{\kappa\rho
}-P_{\sigma\tau}^{\ast}P_{\kappa\rho}^{\ast}\right) \nonumber\\
&  =F_{\sigma\tau}R^{\tau\sigma}-G_{\sigma\tau}S^{\tau\sigma}=2F_{\sigma\tau
}R^{\tau\sigma}\nonumber\\
&  =4\left(  \mathbf{E}\cdot\mathbf{D-H}\cdot\mathbf{B}\right) \nonumber
\end{align}
and%
\begin{align}
\mathcal{L}_{1}  &  =-\mathcal{L}_{1}^{\ast}=P_{\sigma\tau}^{\ast}%
Q^{\tau\sigma}=\frac{1}{2}\left(  P_{\sigma\tau}^{\ast}Q^{\tau\sigma
}-P_{\sigma\tau}\overset{\ast}{\left.  Q^{\tau\sigma}\right.  }\right)
\label{Lagrangian1}\\
&  =\frac{i}{2}e^{\sigma\tau\kappa\rho}P_{\sigma\tau}P_{\kappa\rho}^{\ast
}=4i\left(  \mathbf{E}\cdot\mathbf{B-H}\cdot\mathbf{D}\right)  .\nonumber
\end{align}
Then, by formal differentiation,%
\begin{equation}
\frac{\partial\mathcal{L}_{0}}{\partial P_{\alpha\beta}}=Q^{\beta\alpha
},\qquad\frac{\partial\mathcal{L}_{0}}{\partial P_{\alpha\beta}^{\ast}%
}=\overset{\ast}{\left.  Q^{\beta\alpha}\right.  } \label{DiffZero}%
\end{equation}
and%
\begin{equation}
\frac{\partial\mathcal{L}_{1}}{\partial P_{\alpha\beta}^{\ast}}=Q^{\beta
\alpha},\qquad\frac{\partial\mathcal{L}_{1}^{\ast}}{\partial P_{\alpha\beta}%
}=\overset{\ast}{\left.  Q^{\beta\alpha}\right.  } \label{DiffOne}%
\end{equation}
in view of (\ref{B3}).

The complex covariant Maxwell equations (\ref{CovariantMaxwell}) take the
forms%
\begin{equation}
\frac{\partial}{\partial x^{\nu}}\left(  \frac{\partial\mathcal{L}_{0}%
}{\partial P_{\nu\mu}}\right)  =-\frac{4\pi}{c}j^{\mu},\qquad\frac{\partial
}{\partial x^{\nu}}\left(  \frac{\partial\mathcal{L}_{1}}{\partial P_{\nu\mu}%
}\right)  =\frac{4\pi}{c}j^{\mu} \label{MaxwellLagrangian}%
\end{equation}
and the covariant energy-momentum balance relations
(\ref{CovariantEnergyMomentumBalance}) are given by%
\begin{align}
&  \frac{\partial}{\partial x^{\nu}}\left[  \frac{1}{16\pi}\left(
P_{\mu\lambda}^{\ast}\frac{\partial\mathcal{L}_{0}}{\partial P_{\nu\lambda}%
}+P_{\mu\lambda}\frac{\partial\mathcal{L}_{0}}{\partial P_{\nu\lambda}^{\ast}%
}\right)  \right] \label{EnergyMomentumLagrangian0}\\
&  \quad+\frac{1}{32\pi}\left[  P_{\sigma\tau}^{\ast}\frac{\partial}{\partial
x^{\mu}}\left(  \frac{\partial\mathcal{L}_{0}}{\partial P_{\sigma\tau}%
}\right)  +P_{\sigma\tau}\frac{\partial}{\partial x^{\mu}}\left(
\frac{\partial\mathcal{L}_{0}}{\partial P_{\sigma\tau}^{\ast}}\right)
\right]  =-\frac{1}{c}F_{\mu\lambda}j^{\lambda}\nonumber
\end{align}
and%
\begin{align}
&  \frac{\partial}{\partial x^{\nu}}\left[  \frac{1}{16\pi}\left(
P_{\mu\lambda}\frac{\partial\mathcal{L}_{1}}{\partial P_{\nu\lambda}}%
+P_{\mu\lambda}^{\ast}\frac{\partial\mathcal{L}_{1}^{\ast}}{\partial
P_{\nu\lambda}}\right)  \right] \label{EnergyMomentumLagrangian1}\\
&  \quad+\frac{1}{32\pi}\left[  P_{\sigma\tau}\frac{\partial}{\partial x^{\mu
}}\left(  \frac{\partial\mathcal{L}_{1}}{\partial P_{\sigma\tau}}\right)
+P_{\sigma\tau}^{\ast}\frac{\partial}{\partial x^{\mu}}\left(  \frac
{\partial\mathcal{L}_{1}^{\ast}}{\partial P_{\sigma\tau}^{\ast}}\right)
\right]  =\frac{1}{c}F_{\mu\lambda}j^{\lambda}\nonumber
\end{align}
in terms of the complex Lagrangians under consideration, respectively.

Finally, with the help of the following densities,%
\begin{equation}
L_{0}=\mathcal{L}_{0}-\frac{4\pi}{c}j^{\nu}A_{\nu},\qquad L_{1}=\mathcal{L}%
_{1}+\frac{4\pi}{c}j^{\nu}A_{\nu}, \label{Lagrangians01}%
\end{equation}
one can derive analogs of the Euler-Lagrange equations for electromagnetic
fields in media:%
\begin{equation}
\frac{\partial}{\partial x^{\nu}}\left(  \frac{\partial L_{0,1}}{\partial
P_{\nu\mu}}\right)  -\frac{\partial L_{0,1}}{\partial A_{\mu}}=0.
\label{EulerLagrange}%
\end{equation}
In the case of a moving isotropic medium, a relation between $P_{\nu\mu}$ and
$A_{\mu}$ is given by our equations (\ref{QPmaterialF}%
)--(\ref{4VectorPotential}).

\subsection{Real form}

Taking the real and imaginary parts, Maxwell's equations (\ref{MaxwellII}) can
be written as follows%
\begin{equation}
\partial_{\nu}R^{\mu\nu}=-\frac{4\pi}{c}j^{\mu},\qquad\partial_{\nu}S^{\mu\nu
}=0. \label{MaxwellReal}%
\end{equation}
Here,%
\[
-6\partial_{\nu}S^{\mu\nu}=e^{\mu\nu\lambda\sigma}\left(  \partial_{\nu
}F_{\lambda\sigma}+\partial_{\sigma}F_{\nu\lambda}+\partial_{\lambda}%
F_{\sigma\nu}\right)  \equiv0,
\]
with the help of (\ref{QPIdentity}) and (\ref{4VectorPotential}). Thus the
second set of equations is automatically satisfied when we introduce the
four-vector potential. For the inhomogeneous pair of Maxwell's equations, the
Lagrangian density is given by%
\begin{align}
L  &  =\frac{1}{4}F_{\sigma\tau}R^{\tau\sigma}-\frac{4\pi}{c}j^{\sigma
}A_{\sigma}\label{LagrangianReal}\\
&  =\frac{1}{4}F_{\sigma\tau}\epsilon^{\tau\sigma\lambda\rho}F_{\lambda\rho
}-\frac{4\pi}{c}j^{\sigma}A_{\sigma},\nonumber
\end{align}
in view of (\ref{CovariantMaterialEqs}). Then, for \textquotedblleft conjugate
momenta\textquotedblright\ to the four-potential field $A_{\mu},$ one gets%
\begin{equation}
\frac{\partial L}{\partial\left(  \partial_{\nu}A_{\mu}\right)  }%
=\frac{\partial L}{\partial F_{\sigma\tau}}\frac{\partial F_{\sigma\tau}%
}{\partial\left(  \partial_{\nu}A_{\mu}\right)  }=R^{\mu\nu}
\label{ConjugateMomentum}%
\end{equation}
and the corresponding Euler-Lagrange equations take a familiar form%
\begin{equation}
\partial_{\nu}\left(  \frac{\partial L}{\partial\left(  \partial_{\nu}A_{\mu
}\right)  }\right)  -\frac{\partial L}{\partial A_{\mu}}=0.
\label{EulerLagrangeReal}%
\end{equation}
The latter equation can also be derived with the help of the least action
principle \cite{Pauli}, \cite{TerRyb}, \cite{ToptyginI}. The corresponding
Hamiltonian and quantization are discussed in \cite{HeisenbergQM},
\cite{JauchWatson48a}, \cite{Riazanov57} among other classical accounts.

In conclusion, it is worth noting the role of complex fields in quantum
electrodynamics, quadratic invariants and quantization (see, for instance,
\cite{Acosta-Suslov14}, \cite{Bia:Bia75}, \cite{Bia:Bia13},
\cite{Dodonov:Man'koFIAN87}, \cite{JauchWatson48a}, \cite{JauchWatson48b},
\cite{KlauderSudarshan}, \cite{KretalSus13}, \cite{Lan:Lop:Sus},
\cite{Lop:Sus:VegaGroup}, \cite{LopSusVegaHarm}, \cite{Riazanov57},
\cite{Riazanov58}, \cite{Ryazanov58}, \cite{ToptyginI}, \cite{WatsonJauch49}).
The classical and quantum theory of Cherenkov radiation is reviewed in
\cite{Afanasiev04}, \cite{Bol57}, \cite{Bol09}, \cite{Ginsburg40},
\cite{Ginsburg89}, \cite{Sokolov40}, \cite{Tamm39}. For paraxial approximation
in optics, see \cite{Fock65}, \cite{KisPlach16}, \cite{KoutSuaSus},
\cite{Mah:Sus12}, \cite{Mah:Sus13}, and the references therein. Maxwell's
equations in the gravitational field are discussed in \cite{Carmeli01}
\cite{Fock64}. One may hope that our detailed mathematical consideration of
several aspects of macroscopic electrodynamics will be useful for future
investigations and pedagogy.

\noindent{\textbf{Acknowledgements}}. We dedicate this article to the memory
of Professor Alladi Ramakrishnan who made significant contributions to
probability and statistics, elementary particle physics, cosmic rays and
astrophysics, matrix theory, and the special theory of relativity
\cite{AlladiK}. We are grateful to Krishna Alladi, Albert Boggess,
Mark~P.~Faifman, John~R.~Klauder, Vladimir I. Man'ko, and Igor N.~Toptygin for
valuable comments and help.

\appendix

\section{Formulas from Vector Calculus}

Among useful differential relations are%
\begin{equation}
\nabla\cdot\left(  \mathbf{A}\times\mathbf{B}\right)  =\mathbf{B}\cdot\left(
\nabla\times\mathbf{A}\right)  -\mathbf{A}\cdot\left(  \nabla\times
\mathbf{B}\right)  . \label{A1}%
\end{equation}%
\begin{equation}
\nabla\cdot\left(  f\mathbf{A}\right)  =\left(  \nabla f\right)
\cdot\mathbf{A}+f\left(  \nabla\cdot\mathbf{A}\right)  . \label{A2}%
\end{equation}%
\begin{equation}
\nabla\times\left(  f\mathbf{A}\right)  =\left(  \nabla f\right)
\times\mathbf{A}+f\left(  \nabla\times\mathbf{A}\right)  . \label{A3}%
\end{equation}%
\begin{align}
&  \mathbf{A}\cdot\left(  \nabla\times\left(  f\nabla\times\mathbf{B}\right)
\right)  -\mathbf{B}\cdot\left(  \nabla\times\left(  f\nabla\times
\mathbf{A}\right)  \right) \label{A4}\\
&  \quad=\nabla\cdot\left(  f\left(  \mathbf{B}\times\left(  \nabla
\times\mathbf{A}\right)  -\mathbf{A}\times\left(  \nabla\times\mathbf{B}%
\right)  \right)  \right)  .\nonumber
\end{align}%
\begin{align}
&  \mathbf{A}\left(  \nabla\cdot\mathbf{B}\right)  -\mathbf{B}\left(
\nabla\cdot\mathbf{A}\right)  +\mathbf{A}\times\left(  \nabla\times
\mathbf{B}\right)  -\mathbf{B}\times\left(  \nabla\times\mathbf{A}\right)
\label{A5}\\
&  \quad-\nabla\times\left(  \mathbf{A}\times\mathbf{B}\right)  =\sum
_{\alpha=1}^{3}A_{\alpha}^{2}\nabla\left(  \frac{B_{\alpha}}{A_{\alpha}%
}\right)  =-\sum_{\alpha=1}^{3}B_{\alpha}^{2}\nabla\left(  \frac{A_{\alpha}%
}{B_{\alpha}}\right)  .\nonumber
\end{align}
(See also \cite{Abraham}, \cite{Schwinger} and \cite{ToptyginI}.) Here,
$\operatorname{div}\mathbf{A}=\nabla\cdot\mathbf{A}$ and $\operatorname{curl}%
\mathbf{A}=\nabla\times\mathbf{A}.$

\section{Dual Tensor Identities}

In this article, $e^{\mu\nu\sigma\tau}=-e_{\mu\nu\sigma\tau}$ and
$e_{0123}=+1$ is the Levi-Civita four-symbol \cite{Fock64} with familiar
contractions:%
\begin{equation}
e^{\mu\nu\sigma\tau}e_{\mu\kappa\lambda\rho}=-\left\vert
\begin{array}
[c]{ccc}%
\delta_{\kappa}^{\nu} & \delta_{\lambda}^{\nu} & \delta_{\rho}^{\nu}\\
\delta_{\kappa}^{\sigma} & \delta_{\lambda}^{\sigma} & \delta_{\rho}^{\sigma
}\\
\delta_{\kappa}^{\tau} & \delta_{\lambda}^{\tau} & \delta_{\rho}^{\tau}%
\end{array}
\right\vert , \label{B01}%
\end{equation}%
\begin{equation}
e^{\mu\nu\sigma\tau}e_{\mu\nu\lambda\rho}=-2\left\vert
\begin{array}
[c]{cc}%
\delta_{\lambda}^{\sigma} & \delta_{\rho}^{\sigma}\\
\delta_{\lambda}^{\tau} & \delta_{\rho}^{\tau}%
\end{array}
\right\vert =-2\left(  \delta_{\lambda}^{\sigma}\delta_{\rho}^{\tau}%
-\delta_{\rho}^{\sigma}\delta_{\lambda}^{\tau}\right)  , \label{B02}%
\end{equation}%
\begin{equation}
e^{\mu\nu\sigma\tau}e_{\mu\nu\sigma\rho}=-6\delta_{\rho}^{\tau},\qquad
e^{\mu\nu\sigma\tau}e_{\mu\nu\sigma\rho}=-24. \label{B03}%
\end{equation}

Dual second rank four-tensor identities are given by \cite{Fock64}:%
\begin{equation}
e^{\mu\nu\sigma\tau}A_{\sigma\tau}=2B^{\mu\nu},\qquad e_{\mu\nu\sigma\tau
}B^{\sigma\tau}=A_{\nu\mu}-A_{\mu\nu}. \label{B04}%
\end{equation}
In particular,%
\begin{align}
Q^{\mu\nu}  &  =R^{\mu\nu}+iS^{\mu\nu}=R^{\mu\nu}-\frac{i}{2}e^{\mu\nu
\sigma\tau}F_{\sigma\tau},\label{B1}\\
P_{\mu\nu}  &  =F_{\mu\nu}+iG_{\mu\nu}=F_{\mu\nu}-\frac{i}{2}e_{\mu\nu
\sigma\tau}R^{\sigma\tau}. \label{B2}%
\end{align}%
\begin{equation}
e_{\mu\nu\sigma\tau}Q^{\sigma\tau}=2iP_{\mu\nu},\qquad2iQ^{\mu\nu}=e^{\mu
\nu\sigma\tau}P_{\sigma\tau}. \label{B3}%
\end{equation}%
\begin{equation}
2R^{\mu\nu}=e^{\mu\nu\sigma\tau}G_{\sigma\tau},\qquad-2S^{\mu\nu}=e^{\mu
\nu\sigma\tau}F_{\sigma\tau}. \label{B4}%
\end{equation}%
\begin{equation}
2G_{\mu\nu}=-e_{\mu\nu\sigma\tau}R^{\sigma\tau},\qquad2F_{\mu\nu}=e_{\mu
\nu\sigma\tau}S^{\sigma\tau}. \label{B5}%
\end{equation}%
\begin{equation}
P_{\mu\nu}Q^{\mu\nu}=2F_{\mu\nu}R^{\mu\nu}-\frac{i}{2}\left(  e^{\mu\nu
\sigma\tau}F_{\mu\nu}F_{\sigma\tau}+e_{\mu\nu\sigma\tau}R^{\mu\nu}%
R^{\sigma\tau}\right)  . \label{B6}%
\end{equation}
By direct calculation,%
\begin{equation}
F_{\mu\nu}R^{\mu\nu}=2\left(  \mathbf{H}\cdot\mathbf{B}-\mathbf{E}%
\cdot\mathbf{D}\right)  , \label{B7}%
\end{equation}%
\begin{equation}
e^{\mu\nu\sigma\tau}F_{\mu\nu}F_{\sigma\tau}=8\mathbf{E}\cdot\mathbf{B},\qquad
e_{\mu\nu\sigma\tau}R^{\mu\nu}R^{\sigma\tau}=8\mathbf{H}\cdot\mathbf{D}.
\label{B8}%
\end{equation}
As a result,%
\begin{equation}
\frac{1}{4}P_{\mu\nu}Q^{\mu\nu}=\mathbf{H}\cdot\mathbf{B}-\mathbf{E}%
\cdot\mathbf{D}-i\left(  \mathbf{E}\cdot\mathbf{B}+\mathbf{H}\cdot
\mathbf{D}\right)  . \label{B9}%
\end{equation}
An important decomposition,%
\begin{align}
P_{\mu\lambda}^{\ast}Q^{\lambda\nu}+P_{\mu\lambda}\overset{\ast}{\left.
Q^{\lambda\nu}\right.  }  &  =2\left(  F_{\mu\lambda}R^{\lambda\nu}%
+G_{\mu\lambda}S^{\lambda\nu}\right) \label{B10}\\
&  =4F_{\mu\lambda}R^{\lambda\nu}+\delta_{\mu}^{\nu}F_{\sigma\tau}%
R^{\sigma\tau}\nonumber\\
&  =4F_{\mu\lambda}R^{\lambda\nu}-2\delta_{\mu}^{\nu}\left(  \mathbf{E}%
\cdot\mathbf{D}-\mathbf{H}\cdot\mathbf{B}\right)  ,\nonumber
\end{align}
is complemented by an identity,%
\begin{align}
P_{\mu\lambda}Q^{\lambda\nu}+P_{\mu\lambda}^{\ast}\overset{\ast}{\left.
Q^{\lambda\nu}\right.  }  &  =\frac{1}{4}\left(  P_{\sigma\tau}Q^{\tau\sigma
}+P_{\sigma\tau}^{\ast}\overset{\ast}{\left.  Q^{\tau\sigma}\right.  }\right)
\delta_{\mu}^{\nu}\label{B10a}\\
&  =\frac{1}{2}\left(  \mathbf{E}\cdot\mathbf{D}-\mathbf{H}\cdot
\mathbf{B}\right)  \delta_{\mu}^{\nu}.\nonumber
\end{align}

In matrix form,%
\begin{align}
PQ  &  =\left(  F+iG\right)  \left(  R+iS\right)  =\left(  FR-GS\right)
+i\left(  FS+GR\right)  ,\label{B11}\\
P^{\ast}Q  &  =\left(  F-iG\right)  \left(  R+iS\right)  =\left(
FR+GS\right)  +i\left(  FS-GR\right)  . \label{B12}%
\end{align}
Here,%
\begin{align}
FS  &  =\frac{1}{4}\text{Tr}\left(  FS\right)  I=\left(  \mathbf{E}%
\cdot\mathbf{B}\right)  I,\label{B13}\\
GR  &  =\frac{1}{4}\text{Tr}\left(  GR\right)  I=\left(  \mathbf{H}%
\cdot\mathbf{D}\right)  I. \label{B14}%
\end{align}%
\begin{equation}
FR-GS=\frac{1}{2}\text{Tr}\left(  FR\right)  I=\left(  \mathbf{E}%
\cdot\mathbf{D}-\mathbf{H}\cdot\mathbf{B}\right)  I, \label{B15}%
\end{equation}%
\begin{align}
FR+GS  &  =2FR-\frac{1}{2}\text{Tr}\left(  FR\right)  I\label{B16}\\
&  =2FR-\left(  \mathbf{E}\cdot\mathbf{D}-\mathbf{H}\cdot\mathbf{B}\right)
I.\nonumber
\end{align}%
\begin{equation}
\text{Tr}\left(  FR+GS\right)  =0, \label{B17}%
\end{equation}
where $I=$diag$\left(  1,1,1,1\right)  $ is the identity matrix.

Also,%
\begin{align}
PQ  &  =QP=\left(  \mathbf{F}\cdot\mathbf{G}\right)  I,\label{B18}\\
\det P  &  =\det Q=-\left(  \mathbf{F}\cdot\mathbf{G}\right)  ^{2} \label{B19}%
\end{align}
and%
\begin{align}
\mathbf{F}\cdot\mathbf{G}  &  =\left(  \mathbf{E}+i\mathbf{H}\right)
\cdot\left(  \mathbf{D}+i\mathbf{B}\right) \label{B20}\\
&  =\left(  \mathbf{E}\cdot\mathbf{D}-\mathbf{H}\cdot\mathbf{B}\right)
+i\left(  \mathbf{E}\cdot\mathbf{B}+\mathbf{H}\cdot\mathbf{D}\right)
.\nonumber
\end{align}

Other useful dual four-tensor identities are given by \cite{Fock64}:
\begin{equation}
e^{\mu\nu\sigma\tau}A_{\nu\sigma\tau}=6B^{\mu},\qquad A_{\mu\nu\lambda}%
=e_{\mu\nu\lambda\sigma}B^{\sigma}. \label{B20a}%
\end{equation}
In particular,%
\begin{equation}
6i\frac{\partial Q^{\mu\nu}}{\partial x^{\nu}}=e^{\mu\nu\lambda\sigma}\left(
\frac{\partial P_{\lambda\sigma}}{\partial x^{\nu}}+\frac{\partial
P_{\nu\lambda}}{\partial x^{\sigma}}+\frac{\partial P_{\sigma\nu}}{\partial
x^{\lambda}}\right)  , \label{B21}%
\end{equation}
and%
\begin{equation}
\frac{\partial P_{\mu\nu}}{\partial x^{\lambda}}+\frac{\partial P_{\nu\lambda
}}{\partial x^{\mu}}+\frac{\partial P_{\lambda\mu}}{\partial x^{\nu}}%
=ie_{\mu\nu\lambda\sigma}\frac{\partial Q^{\sigma\tau}}{\partial x^{\tau}}
\label{B22}%
\end{equation}
(see also \cite{KrLanSus15}).

\section{Proof of Identities}

In view of (\ref{PQdual}), or (\ref{B3}), and (\ref{B22}), we can write%
\begin{equation}
\left(  \frac{\partial P_{\mu\nu}}{\partial x^{\lambda}}+\frac{\partial
P_{\nu\lambda}}{\partial x^{\mu}}+\frac{\partial P_{\lambda\mu}}{\partial
x^{\nu}}=ie_{\mu\nu\lambda\sigma}\frac{\partial Q^{\sigma\tau}}{\partial
x^{\tau}}\right)  \overset{\ast}{\left.  Q^{\lambda\nu}\right.  }, \label{C1}%
\end{equation}
or%
\begin{align*}
&  2\overset{\ast}{\left.  Q^{\lambda\nu}\right.  }\frac{\partial P_{\mu\nu}%
}{\partial x^{\lambda}}+\overset{\ast}{\left.  Q^{\lambda\nu}\right.  }%
\frac{\partial P_{\nu\lambda}}{\partial x^{\mu}}\\
&  \quad=i\left(  e_{\mu\nu\lambda\sigma}\overset{\ast}{\left.  Q^{\lambda\nu
}\right.  }\right)  \frac{\partial Q^{\sigma\tau}}{\partial x^{\tau}}%
=-2P_{\mu\sigma}^{\ast}\frac{\partial Q^{\sigma\tau}}{\partial x^{\tau}}%
\end{align*}
by (\ref{B3}). Therefore,%
\begin{equation}
P_{\mu\lambda}^{\ast}\frac{\partial Q^{\lambda\nu}}{\partial x^{\nu}}%
-\frac{\partial P_{\mu\lambda}}{\partial x^{\nu}}\overset{\ast}{\left.
Q^{\lambda\nu}\right.  }=-\frac{1}{2}\overset{\ast}{\left.  Q^{\sigma\tau
}\right.  }\frac{\partial P_{\tau\sigma}}{\partial x^{\mu}}. \label{C2}%
\end{equation}
In addition, with the help of (\ref{B3}) one gets%
\begin{align*}
&  2i\left(  P_{\sigma\tau}^{\ast}\frac{\partial Q^{\tau\sigma}}{\partial
x^{\mu}}\right)  =P_{\sigma\tau}^{\ast}e^{\tau\sigma\lambda\nu}\frac{\partial
P_{\lambda\nu}}{\partial x^{\mu}}\\
&  \quad=e^{\sigma\tau\nu\lambda}P_{\sigma\tau}^{\ast}\frac{\partial
P_{\lambda\nu}}{\partial x^{\mu}}=-2i\left(  \overset{\ast}{\left.
Q^{\sigma\tau}\right.  }\frac{\partial P_{\tau\sigma}}{\partial x^{\mu}%
}\right)  ,
\end{align*}
which completes the proof.

\end{document}